\documentclass[10pt,twocolumn,english,prd,showpacs,nofootinbib]{revtex4}
\usepackage[latin9]{inputenc}
\usepackage{graphicx}
\usepackage{amsmath}
\usepackage{amssymb}
\usepackage{amsfonts}
\usepackage{lscape}
\usepackage{url}
\usepackage{epsfig}
\usepackage{color}
\usepackage{bm}
\usepackage{babel}
\usepackage{xcolor}
\usepackage{tabularx}

\providecommand{\tabularnewline}{\\}


\newcommand{\nn}{\nonumber}
\newcommand{\newc}{\newcommand}
\newc{\be}{\begin{equation}}
\newc{\ee}{\end{equation}}
\newc{\ba}{\begin{eqnarray}}
\newc{\ea}{\end{eqnarray}}
\newc{\bea}{\begin{eqnarray*}}
\newc{\eea}{\end{eqnarray*}}
\def \lcdm {$\Lambda$CDM }
\def \lcdmb {$\Lambda$CDM}
\def \om {\Omega_{\rm m}}

\def \Geff {G_{\rm eff}}

\def \fs8 {$f\sigma_8$ }
\def \OO {$\mathcal{O}$}
\newc{\GB}{\mathcal{G}}

\begin{document}
\title{Reconstruction of the null-test for the matter density perturbations }

\author{Savvas Nesseris$^{1}$}
\email{savvas.nesseris@uam.es}

\author{Domenico Sapone$^{1,2}$}
\email{domenico.sapone@uam.es}

\author{Juan Garc\'ia-Bellido$^{1}$}
\email{juan.garciabellido@uam.es}

\affiliation{$^{1}$Departamento de F\'isica Te\'orica and Instituto de F\'isica Te\'orica, \\
Universidad Aut\'onoma de Madrid IFT-UAM/CSIC,\\ $28049$ Cantoblanco, Madrid, Spain \\ $^{2}$Cosmology and Theoretical Astrophysics group, Departamento de F\'isica, FCFM, Universidad de Chile,\\
Blanco Encalada 2008, Santiago, Chile}


\begin{abstract}
We systematically study the null-test for the growth rate data first presented in \cite{Nesseris:2014mfa} and we reconstruct it using various combinations of data sets, such as the \fs8  and $H(z)$ or Type Ia supernovae (SnIa) data. We perform the reconstruction in two different ways, either by directly binning the data or by fitting various dark energy models. We also examine how well the null-test can be reconstructed by future data by creating mock catalogs based on the cosmological constant model, a model with strong dark energy perturbations, the $f(R)$  and $f(G)$ models, and the large void LTB model that exhibit different evolution of the matter perturbations. We find that with future data similar to an LSST-like survey, the null-test will be able to successfully discriminate between these different cases at the $5\sigma$ level.
\end{abstract}
\maketitle

\section{Introduction}

The late-time accelerated expansion of the Universe has forced cosmologists to revise our understanding of the Universe either by introducing a new component in the Universe called dark energy \cite{sapone-review} or by modifying directly the laws of gravity \cite{tsujikawa}. Within the framework of Friedmann-Lema\^itre-Robertson-Walker (FLRW) cosmology, we can account for a phase of accelerated expansion by simply introducing a cosmological constant ($\Lambda$), even though this model gives rise to severe coincidence and fine-tuning problems, observations are still consistent with a dark energy component that has the same characteristics of the cosmological constant \cite{planck, sdss2014}.

Unfortunately, these experiments are not able to give us information either on the variation on time of a dark energy component or on the {\em clustering properties} of such dark energy component. Moreover, recent observations do not have a sufficient sensitivity to be able to distinguish between a dark energy component or a modified gravity model; even though the two classes of models can be arbitrarily alike, see \cite{Kunz:2006ca}, it is still important to be able to discard some of the model that manifests a different behaviour.

Furthermore, future experiments have been planned to collect a large amount of data with high accuracy and it would be interesting to find tests that are able to confirm our assumptions. Consistency checks are usually model independent tests able to determine if the assumptions made are violated. In this paper we make use of the consistency check found in \cite{Nesseris:2014mfa}. In the latter, we introduced a new null-test specifically for the growth of matter perturbations and, as far as we know, this is the first null-test that accounts for perturbations on the matter fluids. The evolution of the matter density contrast is governed by the evolution of the Hubble parameter and by the evolution of all the other clustering components \cite{saponeDEP, saponeDEP2, Sapone:2012nh, Sapone:2013wda}. Hence it is a complementary probe for the dark energy because, while many different dark energy models give the same expansion history they usually differ at perturbation level (depending on the intrinsic characteristics of the fluid itself) and they will affect the evolution of the matter density contrast.

Moreover, as it is well known, modified gravity models can also be reinterpreted as effective dark energy models with their own {\em effective perturbed quantities} and consequently the growth of matter density will be influenced by these {\em effective} perturbations; see \cite{Kunz:2006ca}. Hence, it was necessary to find a null-test that accounted for the growth of matter density fields.

Finally, the paper is organized as follows: in Sec. \ref{settings} we report the main equations for the growth of matter density contrast; in Sec. \ref{sec:lagrangianform} we review the derivation of the null-test and generalize it to include modified gravity models, in Sec. \ref{sec:nulltest} we construct the null-test and discuss its implications. Finally, in Sec. \ref{sec:datanalysis} we reconstruct the null-test with a variety of data and in Sec. \ref{sec:conclusions} we summarize our results.

\section{Evolution of matter density contrast}\label{settings}

The growth of matter in the Universe under the assumption of homogeneity and isotropy is governed by the second order differential equation:
\ba
\delta''(a)+\left(\frac3a + \frac{H'(a)}{H(a)}\right)\delta'(a)-\frac32 \frac{\Omega_m \Geff(a)/G_N \delta(a)}{a^5 H(a)^2/H_0^2}=0,\nn\\
\label{ode}
\ea
where $H(a)$ is the Hubble parameter, $\Omega_{m_0}$ is the matter density contrast today and $H_0$ is the Hubble constant and where we introduced the effective Newtonian constant  $\Geff (a)$ that accounts for dark energy perturbations or for a variety of modified gravity models; see Refs. \cite{Amendola:2007rr, Tsujikawa:2007gd, Nesseris:2008mq, Nesseris:2009jf}.

Under the assumption that the Universe is currently dominated by a dark energy component with a constant equation of state and negligible dark energy perturbations, i.e. $\Geff(a)/G_N=1$, then Eq.~(\ref{ode}) can be easily solved analytically. The differential equation (\ref{ode}) has in general two solutions that correspond to two different physical modes, a decaying and a growing one, that in a matter dominated Universe in GR behave as $\delta=a^{-3/2}$ and as $\delta=a$ respectively. Since we are only interested in the latter, we demand that at early times $a_{in}\ll1$, usually during matter domination, the initial conditions have to be chosen as $\delta(a_{in})\simeq a_{in}$ and $\delta'(a_{in})\simeq1$. When $\Geff(a)/G_N=1$ we get GR as a subcase, while in general for modified gravity theories, the term $\Geff$ can be scale and time dependent.

For a flat GR model with a constant dark energy equation of state $w$, the exact solution of Eq.~(\ref{ode}) for the growing mode is given by \cite{Belloso:2011ms,Silveira:1994yq, Percival:2005vm}
\ba
\delta(a)&=& a~{}_2F_1\left[- \frac{1}{3 w},\frac{1}{2} -
\frac{1}{2 w};1 - \frac{5}{6 w};a^{-3 w}(1 - \om^{-1})\right]
\label{Da1} \nn \\ \textrm{for}&&H(a)^2/H_0^2= \om a^{-3}+(1-\om)a^{-3(1+w)},\ea
where ${}_2F_1(a,b;c;z)$ is a hypergeometric function, see Ref.~ \cite{handbook} for more details. In more general cases, for instance admitting that the dark energy equation of state parameter is a function of time, it is impossible to find a closed form analytical solution for Eq.~(\ref{ode}), but in Ref.~\cite{Wang:1998gt} it was shown that the growth rate $f(a)\equiv \frac{d ln \delta}{dlna}$ can be approximated as
\ba
f(a)&=&\om(a)^{\gamma(a)} \label{fg}\\
\om(a)&\equiv&\frac{\om~a^{-3}}{H(a)^2/H_0^2} \\
\gamma(a) &=&\frac{\ln f(a)}{\ln \om(a)}\simeq \frac{3 (1-w)}{5-6 w}+\cdots\,
\ea
a more general expression for the growth index can be found in~\cite{Belloso:2011ms}.
We should note that the approximation for $\gamma$ is valid at first order for a dark energy model with a constant $w$, while for \lcdm ($w=-1$) we have $\gamma=\frac{6}{11}\simeq 0.545$. Furthermore, it is easy to convert Eq.~(\ref{ode}) into an equation for the growth rate $f(a)\equiv \frac{d ln \delta}{dlna}$, which can be found to be
\be
f'(a)+\left(\frac{2}{a}+\frac{H'(a)}{H(a)}\right)f(a)+\frac{1}{a}f(a)^2-\frac{3\Omega_{m_0}\Geff(a)/G_N}{2a^4H(a)^2/H_0^2}=0\label{growthode}
\ee
with initial conditions $f(a_0)=1$ for $a_0\ll1$ (usually $a_0\simeq 10^{-3}$).

In later sections we will use the previous equations to construct a null-test for the growth rate of matter density perturbations. The null-test is a function of redshift $z$ that, however, has to be constant for all $z$ under some assumptions, eg that GR is valid or homogeneity holds (since Eq.~(\ref{ode}) was evaluated under the assumption of homogeneity and isotropy). Any deviation from the expected result then indicates the failure of one or more of the assumptions. Typical examples in cosmology include the $\Omega_K(z)$ test of Clarkson et al \cite{Clarkson:2007pz} or the $Om$ statistic of Shafieloo et al., see \cite{Sahni:2008xx}, \cite{Nesseris:2010ep}. Also, early examples of null-tests for the growth data were shown in Refs.~\cite{Nesseris:2007pa} and \cite{Nesseris:2011pc}. However, the former suffers from the problem that we need to know $\delta(z)$ at $z\rightarrow\infty$, while the latter, known as the {\O} test requires some mild assumptions for $\Geff$. On the contrary, we will show that our new null-test does not suffer from any of these problems.

In this paper we expand our work from Ref.~\cite{Nesseris:2014mfa}, creating a null-test that can be used also for more sophisticated cosmological models; we also reconstruct our new null-test with a variety of both real and mock data. The last are created using different cosmologies in order to test the validity and the accuracy of our test.

\section{Lagrangian formulation \label{sec:lagrangianform}}

In this section we will review the derivation of the null-test using the Lagrangian formulation and we will expand it for modified gravity theories. This can be done by again constructing a Lagrangian for Eq.~(\ref{ode}) and with the help of Noether's theorem we can find an associated conserved quantity. If we assume that the Lagrangian can be written as
\be
\mathcal{L}=\mathcal{L}(a,\delta(a),\delta'(a))
\ee
then the Euler-Lagrange equations become:
\be
\frac{\partial \mathcal{L}}{\partial \delta}-\frac{d}{da}\frac{\partial \mathcal{L}}{\partial \delta'}=0\,.
\label{ELeqs}
\ee
We can assume a Lagrangian of the form
\ba
\mathcal{L}&=&T-V \nn \\
T&=&\frac12 f_1(a,H(a))\delta'(a)^2 \nn \\
V&=&\frac12 f_2(a,H(a))\delta(a)^2 \nn
\ea
where the second and third terms are the ``kinetic" and ``potential" terms respectively, and $f_1$ and $f_2$ are two functions that need to be found.
Substituting the last equations in the Euler-Lagrange Eq.~(\ref{ELeqs}), we get
\ba
\delta''(a)&+& \left(\frac{\partial_a f_1(a,H)}{f_1(a,H)}+\frac{H'(a) \partial_H f_1(a,H)}{f_1(a,H)}\right)\delta'(a)\nn\\&+&\frac{f_2(a,H)}{f_1(a,H)}\delta(a)=0\,.
\label{ELeqs1}
\ea
Comparing Eq.~(\ref{ELeqs1}) with Eq.~(\ref{ode}) we immediately find that
\ba
f_1(a,H(a))&=& a^3 H(a)/H_0 \nn \\
f_2(a,H(a))&=& -\frac{3\om \Geff(a)/G_N}{2a^2 H(a)/H_0}\nn \,.
\ea
Then the Lagrangian $\mathcal{L}$ and the `Hamiltonian' $\mathcal{H}$ of the system become:
\ba
\mathcal{L}=T-V&=&\frac12 a^3 H(a)/H_0\delta'(a)^2+\nn\\
&+&\frac{3\om \Geff(a)/G_N}{4a^2 H(a)/H_0}\delta(a)^2 \nn\\ \label{Lag}\\
\mathcal{H}=T+V&=&\frac12 a^3 H(a)/H_0\delta'(a)^2-\nn \\
&-&\frac{3\om \Geff(a)/G_N}{4a^2 H(a)/H_0}\delta(a)^2\,. \label{Ham}\nn\\
\ea
Unfortunately, since the Hamiltonian $\mathcal{H}$ explicitly depends on `time', i.e. the scale factor, the energy of the system is not conserved.

Now that we have obtained the Lagrangian for the system, we can use Noether's theorem to find a conserved quantity that will be later translated into the null-test. So, if we have an infinitesimal transformation $\textbf{X}$ with a generator
\ba
\textbf{X}&=&\alpha(\delta) \frac{\partial }{\partial \delta}+ \frac{d\alpha(\delta)}{da}\frac{\partial }{\partial \delta'} \\
\frac{d\alpha(\delta)}{da}&\equiv&\frac{\partial \alpha}{\partial\delta}\delta'(a)=\alpha'(a),
\ea
such that
\be
L_X \mathcal{L}=0 \label{lieder}
\ee
then
\be
\Sigma=\alpha(a)\frac{\partial \mathcal{L}}{\partial \delta'}\label{const1}
\ee
is a constant of `motion' for the Lagrangian of Eq.~(\ref{Lag}), see Ref.~\cite{Capozziello:2007iu} for an application in Scalar-Tensor cosmology and more details. From Eq.~(\ref{const1}) we get that
\be
\Sigma= a^3 H(a)/H_0 \alpha(\delta)\delta'(a),
\ee
while from Eq.~(\ref{lieder}) we get
\ba
&&\alpha'(a) a^3 H(a)/H_0 \delta'(a)+\nn\\
&+&\frac{3\om \Geff(a)/G_N \delta(a) \alpha(a)}{2a^2H(a)/H_0}=0\,.
\ea
The latter can be solved in favor of $\alpha(a)$ to give
\be
\alpha(a)=c\;e^{-\int_{a_0}^a\frac{3\om  \Geff(x)/G_N\delta(x)}{2x^5 H(x)^2/H_0^2\delta'(x)}dx}
\ee
where $c$ is an integration constant and $a_0$ can be chosen to be either 0 or 1. Then the constraint becomes
\be
\Sigma= a^3 H(a)/H_0 \delta'(a)\;e^{-\int_{a_0}^a\frac{3\om \Geff(x)/G_N \delta(x)}{2x^5 H(x)^2/H_0^2\delta'(x)}dx}\label{const2}
\ee
where we have redefined $\Sigma$ to absorb $c$. Choosing appropriately $a_0$ can lead to convenient values for $\Sigma$, for example for $a_0=1$ then it is easy to see that $\Sigma=\delta'(1)$ and for $a_0\ll1$ then $\Sigma\simeq (\om a_0^3)^{1/2}$, while in general we have $\Sigma=a_0^3 H(a_0)\delta'(a_0)$. We have checked numerically the validity of Eq.~(\ref{const2}) for several different cosmologies and values of the parameters.

Eq.~(\ref{const2}) can also be written in terms of the growth rate $f(a)\equiv \frac{d ln \delta}{dlna}$.
As a consequence, the growth factor can be found to be $\delta(a)=\delta(a_0) e^{\int_{a_0}^a\frac{f(x)}{x}dx}$ and Eq.~(\ref{const2}) can be rewritten as
\be
\Sigma/\delta(a_0)= a^2 H(a) f(a)\; e^{\int_{a_0}^a \left(\frac{f(x)}{x}-\frac{3\om \Geff(x)/G_N}{2x^4 H(x)^2 f(x)}\right)dx}.\label{const1a}
\ee
Taking into account that $\Sigma=a_0^3 H(a_0)\delta'(a_0)$, we get that the LHS of the previous equation
can be re-expressed as
\be
\Sigma/\delta(a_0)=a_0^3 H(a_0)\delta'(a_0)/\delta(a_0)=a_0^2 H(a_0) f(a_0),
\ee
so that Eq.~(\ref{const1a}) becomes:
\be
\frac{a^2 H(a) f(a)}{a_0^2 H(a_0) f(a_0)}\; e^{\int_{a_0}^a \left(\frac{f(x)}{x}-\frac{3\om \Geff(x)/G_N}{2x^4 H(x)^2 f(x)}\right)dx}=1.\label{const2a}
\ee
Taking the derivative of Eq.~(\ref{const2a}) with respect to the scale factor $a$, we obtain Eq.~(\ref{growthode}). This means that Eq.~(\ref{const2a}) is a first integral of ``motion" of Eq.~(\ref{growthode}).

However observations can measure directly only $f\sigma_8(a)\equiv f(a) \sigma_8(a)$,
where $\sigma_8(a)=\sigma_8(a=1)\frac{\delta(a)}{\delta(a=1)}$ and
they are not able to give directly $\delta(a)$,
hence we need to transform Eq.~(\ref{const2}) to be able to test it directly with observations.
Taking into account that
\be
f\sigma_8(a)\equiv f(a) \sigma_8(a)=\xi a \delta'(a),
\ee
where $\xi\equiv \frac{\sigma_8(a=1)}{\delta(a=1)}$, we have that
\be
\delta(a)=\delta(a_0)+\int_{a_0}^a\frac{f \sigma_8(x)}{\xi x}dx.
\ee
Then, Eq.~(\ref{const2}) can be written as
\ba
& &\frac{a^2 H(a) f\sigma_8(a)}{a_0^2 H(a_0)f\sigma_8(a_0)} \cdot \nn\\
& & e^{-\frac32\Omega_m\int_{a_0}^a\frac{\Geff(x)}{G_N}\frac{\sigma_8(a=1)\frac{\delta(a_0)}{\delta(1)}+\int_{a_0}^x\frac{f\sigma_8(y)}{y}dy}{x^4H(x)^2/H_0^2f\sigma_8(x)}dx}=1 \,.
\label{nullf}
\ea
It is clear that the expressions of Eqs.~(\ref{const2a})-(\ref{nullf}) have to be constant for all redshifts $z$, so in the next Section we will use them to construct a null-test. Any deviation from unity will imply the presence of new physics or systematics in the data.

\section{The null-test \label{sec:nulltest}}
In this Section we will use Eqs.~(\ref{const2a})-(\ref{nullf}) and assume $\Geff/G_N=1$ to construct a new null-test for the growth data. Since this equation only holds for GR with the FLRW metric, deviations point to either new physics or systematics in the data. We have explicitly tested in the case of $w=\textrm{const}.$, where the analytical solution is known, that Eqs.~(\ref{const2a}) - (\ref{nullf}) are valid at all redshifts.

In order to create our null-test, we implement Eqs.~(\ref{const2a}) - (\ref{nullf}).
We now have two equivalent forms of the null-test:
\ba
\mathcal{O}(z)&=&\frac{a^2 H(a) f(a)}{a_0^2 H(a_0) f(a_0)}\; e^{\int_{a_0}^a \left(\frac{f(x)}{x}-\frac{3\om}{2x^4 H(x)^2 f(x)}\right)dx} \label{nulltest1}\\
\mathcal{O}(z)&=&\frac{a^2 H(a) f\sigma_8(a)}{a_0^2 H(a_0) f\sigma_8(a_0)}\times\nn\\
&\times&e^{-\frac32\Omega_m\int_{a_0}^a\frac{\sigma_8(a=1)\frac{\delta(a_0)}{\delta(1)}+\int_{a_0}^x\frac{f\sigma_8(y)}{y}dy}{x^4H(x)^2/H_0^2f\sigma_8(x)}dx} \,.
\label{nulltest2}
\ea
Both forms are totally equivalent: Eq.~(\ref{nulltest1}) is expressed in terms of the growth rate $f(a)$, which is not a direct measurable quantity but it makes the expression for the null-test much simpler and it will be useful (as it will be clear later on) for testing directly specific models; Eq.~(\ref{nulltest2}) is written in terms of direct measurable quantities and it will be extremely useful to test the data. It is clear that in both cases
we should have $\mathcal{O}(z)=1$ at all redshifts, and any deviation from unity could be due to several reasons:
\begin{itemize}
  \item Detection of modified gravity and nonconstant $\Geff$.
  \item New physics or a presence of shear or strong dark energy perturbations.
  \item Deviation from the FLRW metric and homogeneity.
  \item Tension between $H(z)$ (obtained directly or derived) and $f\sigma_8$ data.
\end{itemize}

In the next Section we will test the above expression with the help of mocks based on different models.

\section{Cosmological models\label{theory}}

In this section we schematically report the different cosmologies used to create mocks catalogs.

\subsection{$w$CDM model}

If the Universe is filled by a dark energy component, with constant equation of state parameter $w$,
then the Hubble equation can be written as:
\be
H(a)^2/H_0^2= \om a^{-3}+(1-\om)a^{-3(1+w)}\,.
\ee
If the dark energy component is not a cosmological constant, i.e. if $w\neq -1$, then dark energy is able to cluster.
The scale at which this dark energy component can cluster depends on the intrinsic characteristic of the fluid itself, namely: pressure perturbations $\delta p$ which is related to the sound speed and anisotropic stress $\sigma$ that are usually related to the viscosity of the fluid (see \cite{saponeDEP, saponeDEP2, Sapone:2012nh, Sapone:2013wda}). If dark energy is able to cluster at sufficiently small scales then Eq.~(\ref{ode}) needs to be modified to account for the dark energy perturbations which will be an extra source term to the gravitational potential. Since Eq.~(\ref{ode}) has been evaluated on the limit of small scales, dark energy has to have a small value of the sound speed and zero viscosity term. In order to create mock catalogs for the $f\sigma_8(a)$ we modify Eq.~(\ref{ode}) by using the function
\be
Q(a) = 1+ \frac{1-\Omega_{m_0}}{\Omega_{m_0}}\frac{1+w}{1-3w}a^{-3w}
\label{eq:q}
\ee
which will act as a modified Newton's constant, i.e. $\Geff(a)/G_N \equiv Q(a)$. Eq.~(\ref{eq:q}) describes the amount of the dark energy perturbations and it has been evaluated under the assumption of zero dark energy sound speed and zero anisotropic stress, \cite{saponeDEP}. If we set $w=-1$ then we recover the $\Lambda$CDM model, i.e. with zero perturbations. In this case our default parameters for the mocks are: $(\om,w,\sigma_8)=(0.3,-0.8,0.8)$.

\subsection{$f(R)$ model}
Since we are interested in examining the effect of the \fs8 data on the null-test, we choose an $f(R)$ model that is exactly \lcdm at the background level, but is significantly different at the perturbations level. This way, we can disentangle the effects of the modified gravity, $\Geff(a)$ from the background acceleration. One such degenerate model was studied in Ref.~\cite{Nesseris:2013fca}, where the $f(R)$ action was found to be:
\be
S= \frac{1}{8 \pi G_N} \int d^4 x \sqrt{-g}\left(f(R)/2+S_m\right), \label{actionfR}
\ee
where $S_m$ is the action term for the matter fields and
\ba
f(R)&=&R-2\Lambda+\alpha H_0^2\left(\frac{\Lambda }{R-3 \Lambda }\right)^{b} \nn\\ && {}_2F_1\left(b,\frac{3}{2}+b,\frac{13}{6}+2b,\frac{\Lambda }{R-3 \Lambda }\right),\label{fRmodact1}
\ea
where $b=\frac{1}{12} \left(-7+\sqrt{73}\right)$ and the parameter $\alpha$ is dimensionless and determines how strong the effects of the modified gravity are. We should note that with this Lagrangian we can recover GR at early times ($a\ll1$), i.e. $\Geff / G_N\sim 1$ or $f'(R)\sim 1$ and that it passes all criteria for viability of $f(R)$ models, as shown in Ref.~\cite{Nesseris:2013fca}. For this model we have by construction
\be
H(a)^2=H_0^2\left(\om a^{-3}+1-\om\right), \label{HaGR}
\ee
while Newton's constant is \cite{Tsujikawa:2007gd}:
\ba
\Geff / G_N&=&\frac{1}{F}\frac{1+4\frac{k^2}{a^2}m}{1+3\frac{k^2}{a^2}m}, \label{GefffR}\\ m&\equiv& \frac{F_{,R}}{F},\\F&\equiv&f_{,R}=\frac{\partial f}{\partial R}.
\ea
In this case our default parameters for the mocks are: $(\om,\sigma_8)=(0.3,0.8)$ and we also considered the two different cases $\alpha=(0.002,0.2)$.

\subsection{Gauss-Bonnet model}
Another interesting case are the $f(\GB)$ models, where $G$ is the Gauss-Bonnet term $\GB\equiv R^2-4 R_{\mu\nu}R^{\mu\nu}+R_{\mu\nu \sigma\rho}R^{\mu\nu \sigma\rho}$. Again, we are primarily interested in the effects of the modification of gravity, so we will use the $f(G)$ degenerate model of Ref.~\cite{Nesseris:2013fca}, that is exactly \lcdm at the background level. Then the action is given by \cite{Nesseris:2013fca}:
\be
S= \frac{1}{8 \pi G_N} \int d^4 x \sqrt{-g}\left(R/2+f(\GB)\right)+S_m, \label{action}
\ee
where
\be
f(\GB)=-3H_0^2 (1-\om)+\alpha~H_0^2 \GB \int \frac{a(\GB) H(\GB)/H_0}{\GB^2} d\GB . \label{myaction1}
\ee
In the last equation the first term corresponds to the cosmological constant, we have neglected a term that was just proportional to $\GB$ as it does not contribute in the field equations. The cosmological perturbations of the $f(\GB)$ models were studied in Ref.~\cite{DeFelice:2009rw}, where it was shown that the growth factor for the matter perturbations $\delta_m$ satisfies the evolution equation (using the subhorizon approximation $k\gg aH$):
\be
\label{per1}
\ddot{\delta}_m+C_1(k,a) \dot{\delta}_m+C_2(k,a) \delta_m\simeq 0,
\ee
where the functions $C_1(k,a)$ and $C_2(k,a)$ where first derived in \cite{DeFelice:2009rw} and are given in Appendix B of Ref.~\cite{Nesseris:2013fca} for completeness. In the GR limit Eq.~(\ref{per1}) reduces to \be \ddot{\delta}_m+2 H \dot{\delta}_m-\frac{3}{2} \om a^{-3} \delta_m= 0 \ee so comparing these two expressions we can define an effective  Newton's constant:
\be
\Geff(k,a)/G_N= \frac{C_2(k,a)}{-\frac{3}{2} \om a^{-3}}, \label{geff}
\ee
which is valid only under the subhorizon approximation $k\gg aH$.

Even though these models suffer from instabilities in the matter density perturbations during the matter era as shown in \cite{DeFelice:2009rw}, we still use them to make mocks since they exhibit rich phenomenology due to the presence of the second term containing $C_1(k,a)$ in Eq.~(\ref{per1}). This makes them ideal candidates for our null-test, as $C_1(k,a)$ cannot be described by a single $\Geff$ term, thus will produce deviations from unity. In this case our default parameters for the mocks are: $(\om,\alpha,\sigma_8)=(0.3,0.02,0.8)$.

\subsection{LTB model}

Alternatives to $\Lambda$ for explaining the current acceleration are inhomogeneous universe models in which the effective acceleration is caused by our special position as observers inside a huge underdense region of space. One of the simplest models to study the effect of such large inhomogeneities is the spherically symmetric Lema\^{i}tre-Tolman-Bondi model \cite{lemaitre, tolman, bondi} (LTB). In this large void model, the metric is given by
\be
ds^2 = - dt^2 + X^2(r,t)\,dr^2 + A^2(r,t)\,d\Omega^2\,,
\ee
where $d\Omega^2 = d\theta^2 + \sin^2\theta d\phi^2$ and the equivalent of the scale factor now depends on both time and the radial coordinate $r$. We can find a relationship between $X(r,t)$ and $A(r,t)$ using the the $0-r$ component of the Einstein equations: $X(r,t)=A'(r,t)/\sqrt{1-k(r)}$ where a prime denotes a derivative with respect to coordinate $r$ and $k(r)$ is an arbitrary function that plays the role of the spatial curvature parameter.

To find the growth index in LTB cosmologies we must study linear perturbation theory in inhomogeneous Universes. Due to the loss of a degree of symmetry, the decomposition theorem does no longer hold. This means that, in general, our perturbations will no longer decouple into scalar, vector and tensor modes. A study of the perturbation equations in this scenario using a 1+1+2 decomposition of spacetime can be found in \cite{clarkson}. However, if the normalized shear $\varepsilon=(H_T-H_L)/(2H_T+H_L)$ is
small,~\footnote{We now have two different expansion rates, $H_T(r,t)=\dot A/A$ and $H_L(r,t)=\dot A'/A'$, corresponding to the transverse and along the line of sight expansion rates,  respectively.} as observations seem to confirm \cite{shear,Alonso:2012ds}, we can use the ADM formalism and express our perturbed LTB metric as
\be
ds^2 = -(1 + 2\Phi)dt^2+ (1-2\Psi)\gamma_{ij}dx^idx^j\,,
\ee
where $\gamma_{ij}=\text{diag}\{X^2(r,t),A^2(r,t),A^2(r,t)\sin^2\theta\}$. Within this formalism, the growing mode of the density contrast is given by~\cite{Alonso:2012ds}
\be
\delta(r,t)=\frac{A(r,t)}{r} \ \!{}_2\!F_1\Big[1, 2, \frac{7}{2}; u\Big]\,,
\ee
where $u = k(r)A(r, t)/F(r)$ and $F (r) = H_0^2(r) \Omega_M(r) r^3$ specifies the local matter density today.

We can now calculate the growth rate of density perturbations, noting that here the matter density parameter $\Omega_M(r)$ is a function of redshift via both time $t$ and the radial coordinate $r$. In LTB models, this is in principle an arbitrary function which must be chosen appropriately in each case. In the case of the constrained GBH model \cite{GBH,kSZ} the parameters are given by
\begin{eqnarray}
\Omega_M(r) &\!=\!& 1 + (\Omega_M^{(0)}-1){1-\tanh[(r-r_0)/2\Delta r]\over1+\tanh[r_0/\Delta r]}\\
H_0(r) &\!=\!& H_0\left[{1\over1-\Omega_M(r)} - {\Omega_M(r)\over(1-\Omega_M(r))^{3/2}}\times \right. \nonumber\\
&& \hspace{9mm} \left.\times{\rm arcsinh}\sqrt{1-\Omega_M(r)\over\Omega_M(r)}\right]\,,
\end{eqnarray}
with
\begin{equation}
r_0 = 3.0\ {\rm Gpc}\,, \hspace{1mm} \Delta r = \,r_0\,,  \hspace{1mm}
h_0 = 0.71\,,  \hspace{1mm} \Omega_M^{(0)} = 0.19\,,
\end{equation}
where these values have been chosen to best fit the supernovae and BAO data \cite{shear,Zumalacarregui:2012pq}.
Within this model, the growth function, i.e. the logarithmic derivative of the density contrast, is given by~\cite{Belloso:2011ms}
\be\label{fzLTB}
f(z)=\Omega_m^{1/2}(z)\frac{P_{-1/2}^{-5/2}\Big[\Omega_m^{-1/2}(z)\Big]}{P_{1/2}^{-5/2}\Big[\Omega_m^{-1/2}(z)\Big]}\,,
\ee
where $P_l^m(u)$ are the associated Legendre polynomials and $\Omega_m(z)$ is the fraction of matter density to critical density, as a function of redshift.\footnote{The matter density in LTB model is given by $\rho(r,t)=F'(r)/A'(r,t)A^2(r,t)$. Note that this is different from $\Omega_M(r)=F(r)/A^3(r,t_0)H_0^2(r)$, which gives the mass radial function today, see Ref.~\cite{GBH}.} This function (\ref{fzLTB}) is identical to the {\em instantaneous} growth function of matter density in an open Universe, where the local matter density $\Omega_M$ is given by $\Omega_m(z)$ at that redshift. This is a good approximation only in LTB models with small cosmic shear,  see Ref.~\cite{AGBHV}.

\begin{table}
\begin{centering}\begin{tabular}{cccc}
\hline
Index & $z$ & $f\sigma_8(z)$ & Refs. \tabularnewline
\hline
1 & $0.02$ & $0.360\pm 0.040$ & \cite{hudson} \tabularnewline
\hline
2 & $0.067$ & $0.423\pm 0.055$ & \cite{beutler} \tabularnewline
\hline
3 & $0.25$ & $0.3512\pm 0.0583$ & \cite{samushia} \tabularnewline
\hline
4 & $0.37$ & $0.4602\pm 0.0378$ & \cite{samushia} \tabularnewline
\hline
5 & $0.30$ & $0.407\pm 0.055$ & \cite{tojeiro} \tabularnewline
\hline
6 & $0.40$ & $0.419\pm 0.041$ & \cite{tojeiro} \tabularnewline
\hline
7 & $0.50$ & $0.427\pm 0.043$ & \cite{tojeiro} \tabularnewline
\hline
8 & $0.60$ & $0.433\pm 0.067$ & \cite{tojeiro} \tabularnewline
\hline
9 & $0.17$ & $0.510\pm 0.060$ & \cite{song} \tabularnewline
\hline
10 & $0.35$ & $0.440\pm 0.050$ & \cite{song} \tabularnewline
\hline
11 & $0.77$ & $0.490\pm 0.018$ & \cite{song, guzzo} \tabularnewline
\hline
12 & $0.44$ & $0.413\pm 0.080$ & \cite{blake} \tabularnewline
\hline
13 & $0.60$ & $0.390\pm 0.063$ & \cite{blake} \tabularnewline
\hline
14 & $0.73$ & $0.437\pm 0.072$ & \cite{blake} \tabularnewline
\hline
15 & $0.80$ & $0.470\pm 0.080$ & \cite{delatorre} \tabularnewline
\hline
16 & $0.35$ & $0.445\pm 0.097$ & \cite{chuang} \tabularnewline
\hline
17 & $0.32$ & $0.384\pm 0.095$ & \cite{chuangetal} \tabularnewline
\hline
18 & $0.57$ & $0.423\pm 0.052$ & \cite{samushia2} \tabularnewline
\hline
\end{tabular}\par\end{centering}
\caption{$f\sigma_8(z)$ measurements from different surveys.
\label{tab:fs8-data}}
\end{table}

\section{Data analysis\label{sec:datanalysis}}

In this section we present the analysis implemented for the null-test and we describe the data we used.

\subsection{Binning the data}

We reconstruct the null-test $\mathcal{O}(z)$ by using different cosmological measurements.
In order to reconstruct Eq.~(\ref{nullf}) we need four independent observables: the Hubble parameter $H(z)$, the $f\sigma_8(z)$, $\sigma_8(z=0)$ and $\Omega_{m_0}h^2$. To be more specific, we use the $f\sigma_8$ data from different experiments and collected by Refs \cite{Basilakos:2013nfa} and \cite{taddei}, and we reported them in Table \ref{tab:fs8-data}, and the Hubble parameters values measured from passively evolving galaxies data given in Moresco et al. \cite{moresco_etal} and the values of the Hubble parameters using radial Baryon Acoustic Oscillation (BAO) from different experiments \cite{Gaztanaga:2008xz, Blake:2012pj} and \cite{Font-Ribera:2014wya}, we report the values in Table~\ref{tab:h-data}.

\begin{table}
\begin{centering}\begin{tabular}{cccc}
\hline
Index & $z$ & $H(z)$ & Refs. \tabularnewline
\hline
1 & $0.090$ & $69\pm 12$ & \cite{moresco_etal} \tabularnewline
\hline
2 & $0.170$ & $83\pm 8$ &\cite{moresco_etal} \tabularnewline
\hline
3 & $0.179$ & $75\pm 4$ &\cite{moresco_etal} \tabularnewline
\hline
4 & $0.199$ & $75\pm 5$ &\cite{moresco_etal} \tabularnewline
\hline
5 & $0.270$ & $77\pm 14$ &\cite{moresco_etal} \tabularnewline
\hline
6 & $0.352$ & $83\pm 14$ &\cite{moresco_etal} \tabularnewline
\hline
7 & $0.400$ & $95\pm 17$ &\cite{moresco_etal} \tabularnewline
\hline
8 & $0.480$ & $97\pm 62$ & \cite{moresco_etal} \tabularnewline
\hline
9 & $0.593$ & $104\pm 13$ & \cite{song} \tabularnewline
\hline
10 & $0.680$ & $92\pm 8$ & \cite{moresco_etal} \tabularnewline
\hline
11 & $0.781$ & $105\pm 12$ & \cite{moresco_etal} \tabularnewline
\hline
12 & $0.875$ & $125\pm 17$ & \cite{moresco_etal} \tabularnewline
\hline
13 & $0.880$ & $90\pm 40$ & \cite{moresco_etal} \tabularnewline
\hline
14 & $0.900$ & $117\pm 23$ & \cite{moresco_etal} \tabularnewline
\hline
15 & $1.037$ & $154\pm 20$ & \cite{moresco_etal} \tabularnewline
\hline
16 & $1.300$ & $168\pm 17$ & \cite{moresco_etal} \tabularnewline
\hline
17 & $1.430$ & $177\pm 18$ &\cite{moresco_etal} \tabularnewline
\hline
18 & $1.530$ & $140\pm 14$ & \cite{moresco_etal} \tabularnewline
\hline
19 & $1.750$ & $202\pm 40$ & \cite{moresco_etal} \tabularnewline
\hline
20 & $0.240$ & $79.69\pm 2.32$ & \cite{Gaztanaga:2008xz} \tabularnewline
\hline
21 & $0.430$ & $86.45\pm 3.27$ & \cite{Gaztanaga:2008xz} \tabularnewline
\hline
22 & $0.440$ & $82.60\pm 7.80$ &  \cite{Blake:2012pj} \tabularnewline
\hline
23 & $0.570$ & $96.80\pm 3.40$ & \cite{Anderson:2013zyy} \tabularnewline
\hline
24 & $0.600$ & $87.90\pm 6.10$ &  \cite{Blake:2012pj} \tabularnewline
\hline
25 & $0.730$ & $97.30\pm 7.00$ &  \cite{Blake:2012pj} \tabularnewline
\hline
26 & $2.36$ & $226.0\pm 8.00$ & \cite{Font-Ribera:2014wya} \tabularnewline
\hline
\end{tabular}\par\end{centering}
\caption{$H(z)$ measurements from different surveys using passively evolving galaxies and radial BAO.
\label{tab:h-data}}
\end{table}

The binning technique to measure \OO$(z)$ consists of evaluating it in several redshift bins by directly computing the $H(z)$ values and by using the $f\sigma_{8}(z)$ values measured by different experiments. The Hubble parameter catalog contains $n_{H}=26$ data spanned from redshift $0.1$ up to $z=2.36$, whereas the growth rate catalog contains $n_{f\sigma_8}=18$ data points from $z=0.02$ up to $z=0.8$. Since the growth measurements reach only up to  $z=0.8$ we are forced to discard the last 9 data points for $H(z)$ (as we want to avoid having too wide bins). Because the number of data for both catalogs is quite small the choice of the bins is quite restricted. We decided to opt for two different binning: first we chose $4$ and then $3$ bins, both equally spaced and we evaluate the observables at the mean redshift of the bins.

\begin{figure}
\centering
\vspace{0cm}\rotatebox{0}{\vspace{0cm}\hspace{0cm}\resizebox{0.46\textwidth}{!}{\includegraphics{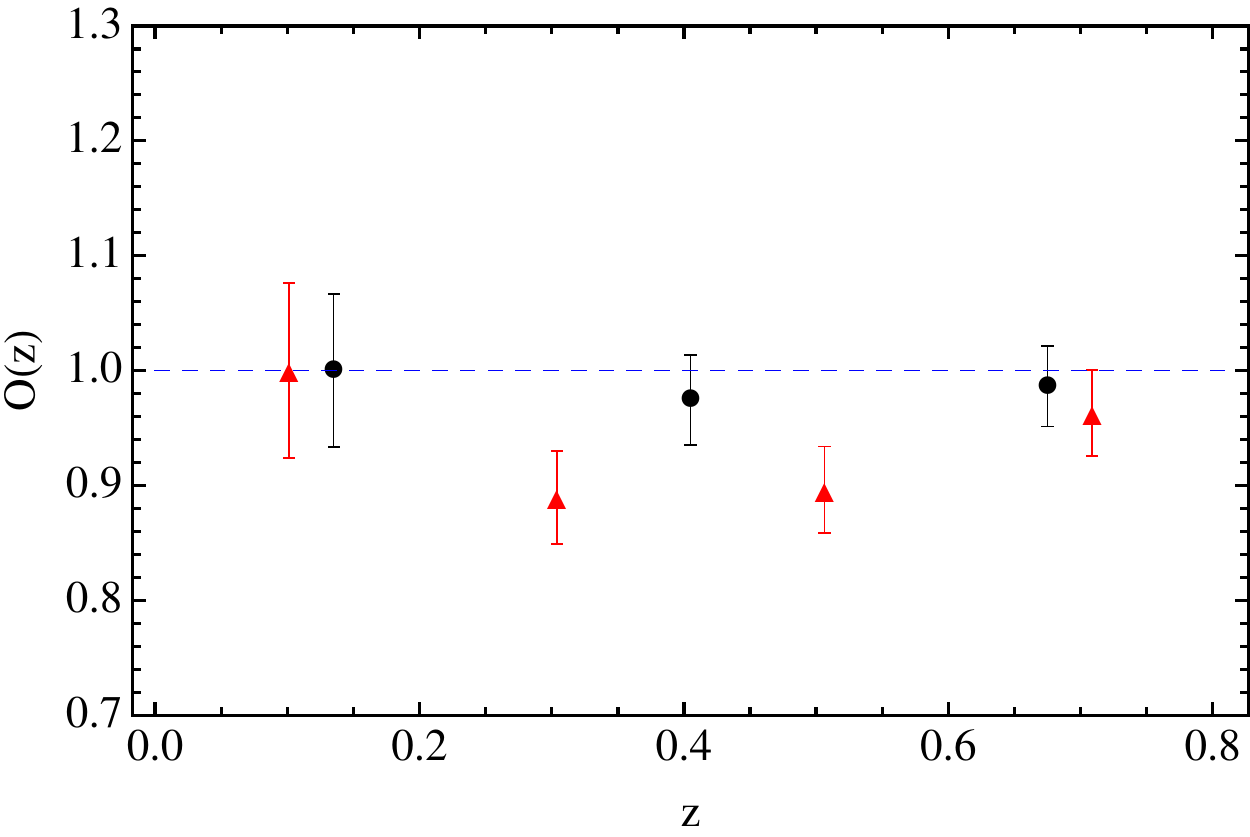}}}
\caption{The results for the null-test with the binning method for the actual data with 4 bins (red triangle) and three bins (black circle), using $H(z)$ and $f\sigma_8(z)$.
\label{fig:binrealvsmock-null}}
\end{figure}

It is important to notice that, in order for the consistency check $\mathcal{O}$ to hold, we need to evaluate quantities at the same redshift. We show the results in Fig.~\ref{fig:binrealvsmock-null}. As can be seen from the figure, the number of bins affect the results; in the $4$-bins case the null-test $\mathcal{O}$ is far from unity implying that the actual data do not give a \lcdm scenario as at redshift $\sim 0.5$ the reference cosmology is at almost $3\sigma$'s away. However, in the $3$-bins case the data predict a \lcdm scenario already at $1\sigma$. The reason why we have such different results is due to the number of data points we are considering. At the moment, we have few data especially for the growth factor and also not uniformly distributed, leaving some bins with only two points and making the binning technique not fully reliable.

\subsection{Mock catalogs}

As mentioned before, we also use mock catalogs based on different cosmologies to test $\mathcal{O}(z)$ for two main reasons:
first, to evaluate how much the errors on the null-test will be with future experiments; second, to examine the validity and the generality of the null-test $\mathcal{O}(z)$.

We used different cosmologies to evaluate the mock catalogs: 1) $w$CDM with $w=-1$ to recover the $\Lambda$CDM limit and another set of data using $w=-0.8$ which allows perturbations in the dark energy sector; 2) $f(R)$ model with $(\om,\sigma_8)=(0.3,0.8)$ and we also considered the two different cases $\alpha=(0.002,0.2)$; 3) $f(G)$ model with $(\om,\alpha,\sigma_8)=(0.3,0.02,0.8)$; and 4) $LTB$ model with $(r_0, \Delta r, r_0, h_0, \Omega_M^{(0)}) = (3.0 {\rm Gpc}, 3.0 {\rm Gpc}, 0.71, 0.19)$. The details of the models can be found in Section \ref{theory}. We created two different catalogs for each cosmology: the Hubble parameter and the $f\sigma_8(z)$. Since we are more interested in testing the consistency check $\mathcal{O}(z)$ rather than worrying about systematics in the data, we evaluated the Hubble and growth parameters uniformly distributed in the range $z \in [0, 2]$ divided into $20$ equally spaced bins of step d$z=0.1$;
the $H(z)$ and the $f\sigma_8(z)$ were estimated as its theoretical value plus a gaussian error (that can be negative or positive) and constant errors of 0.2 and 0.006, respectively;
the values of the errors were obtained using the Fisher matrix approach and having in mind a setup similar to Euclid-like and LSST-like surveys \cite{Amendola:2012ys, Abate:2012za}, i.e. evaluating the sensitivity that future survey will have to measure the Hubble parameter and the growth of matter.

\begin{table}
\begin{centering}
\begin{tabular}{|c|c|c|c|c|c|}
\hline
\multicolumn{6}{|c|}{DATA} \tabularnewline
\hline
 \multicolumn{3}{|c|}{4 bins} & \multicolumn{3}{c|}{3 bins} \tabularnewline
\hline
$\,\,\,z\,\,\,$ & \OO$(z) \pm \sigma_{\mathcal{O}(z)}$ & $\sigma$'s & $z$ &\OO$(z) \pm \sigma_{\mathcal{O}(z)}$ & $\sigma$'s\tabularnewline
\hline
 $0.101$ & $1.000 \pm 0.076$ &  $0.$ & $0.135$ & $1.000\pm 0.066$ & $0.$  \tabularnewline
\hline
 $0.304$ & $ 0.889 \pm 0.041 $ & $2.740$ &$0.405$ & $0.974\pm 0.039$ &$0.656$\tabularnewline
\hline
 $0.506$ & $ 0.896 \pm 0.038 $  & $2.744$ &$0.675$ &$0.986\pm 0.035$ &$0.395$ \tabularnewline
\hline
 $0.709$ & $ 0.963 \pm 0.037 $  & $0.9901$ & ... & ...&  ... \tabularnewline
\hline
\end{tabular}\par\end{centering}
\caption{Null-test \OO$(z)$ with the corresponding $1\sigma$ errors using actual data divided in 4 and 3 bins with the corresponding confidence level.
\label{tab:Nulltest-results-binning}}
\end{table}

\subsection{Binning the mock catalogs}

\begin{table*}
\begin{centering}
\begin{tabular}{|c|c|c|c|c|c|c|c|c|c|c|}
\hline
\multicolumn{11}{|c|}{MOCK} \tabularnewline
\hline
 &  \multicolumn{2}{c|}{\lcdm}  &  \multicolumn{2}{c|}{$w$CDM}  &  \multicolumn{2}{c|}{$f(R)$}&  \multicolumn{2}{c|}{$f(G)$}&  \multicolumn{2}{c|}{LTB}\tabularnewline
\hline
z  & \OO$(z) \pm \sigma_{\mathcal{O}(z)}$  & $\sigma$'s& \OO$(z) \pm \sigma_{\mathcal{O}(z)}$  & $\sigma$'s& \OO$(z) \pm \sigma_{\mathcal{O}(z)}$  & $\sigma$'s& \OO$(z) \pm \sigma_{\mathcal{O}(z)}$  & $\sigma$'s& \OO$(z) \pm \sigma_{\mathcal{O}(z)}$  & $\sigma$'s \tabularnewline
\hline
$0.1$& $1.000\pm 0.010$ & $0$ & $1.000\pm 0.010$ & $0$& $1.000\pm 0.010$& $0$ & $1.000\pm 0.009$& $0$ & $1.000 \pm 0.010$ & $0$\tabularnewline
\hline
$0.3$ & $ 0.994\pm 0.013$ & $0.447$ & $0.971 \pm 0.013$ & $2.302$& $0.975\pm 0.013$ & $1.924$ & $0.983\pm 0.012$ &$1.428$ &$1.327\pm 0.035$ & $9.191$\tabularnewline
\hline
$0.5$ & $0.989 \pm 0.022$ &$0.498$ &$0.970 \pm 0.021$ & $1.421$ & $0.978\pm 0.022$ & $0.987$ & $0.958\pm 0.020$ &$2.154$ &$1.612\pm0.085$ &$7.250$ \tabularnewline
\hline
$0.7$ & $0.979 \pm 0.032$ &$0.637$ &$0.962\pm 0.031$ & $1.203$ & $0.964\pm 0.032$ & $1.111$ &$0.885 \pm 0.027$ & $4.205$ & $1.941\pm0.153$& $6.151$ \tabularnewline
\hline
\end{tabular}\par\end{centering}
\caption{Null-test \OO$(z)$ with $1\sigma$ errors for the five cosmologies used in this work. We also show the confidence level for each test at each redshifts, values less then $1$ indicate that the null-test is consistence with unity at $1\sigma$, if it is larger it corresponds to the sigmas away the null-test is.
\label{tab:Nulltest-results-binning-mocks}}
\end{table*}

\begin{figure*}
\centering
\vspace{0cm}\rotatebox{0}{\vspace{0cm}\hspace{0cm}\resizebox{0.49\textwidth}{!}{\includegraphics{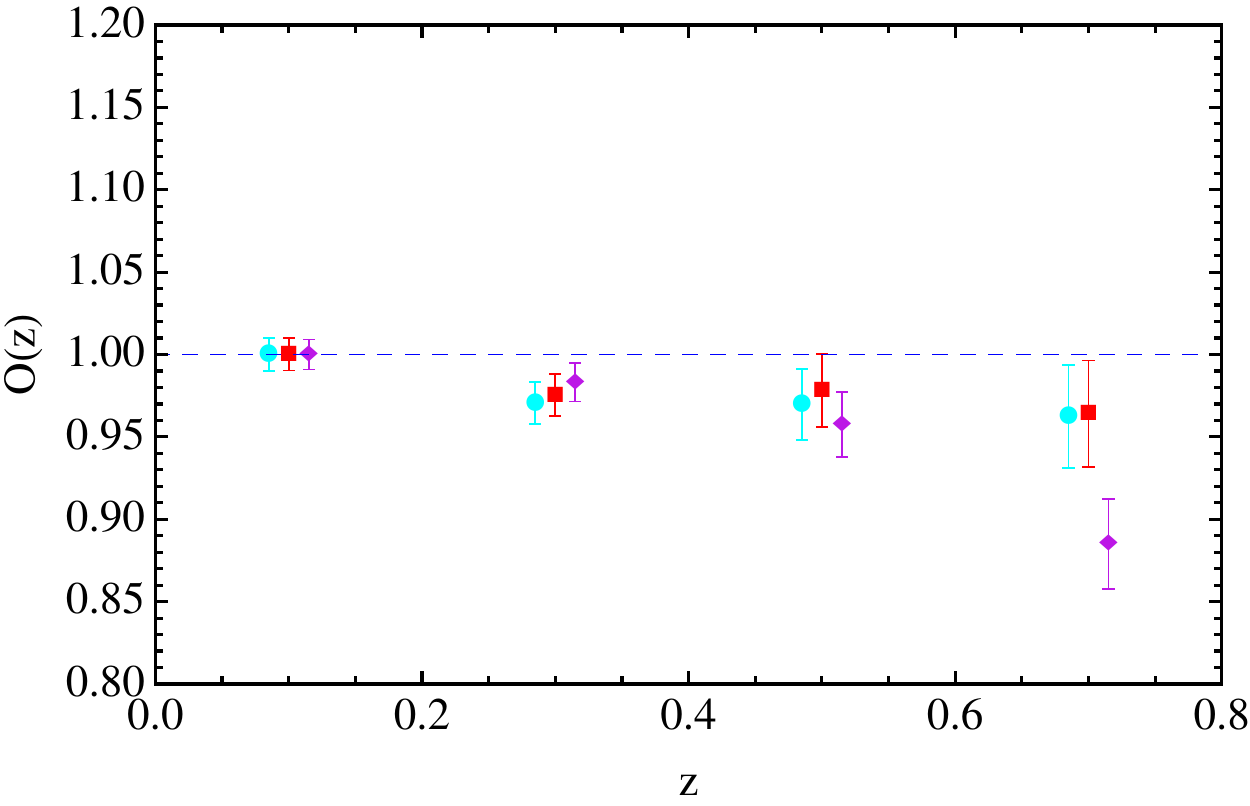}}}
\vspace{0cm}\rotatebox{0}{\vspace{0cm}\hspace{0cm}\resizebox{0.48\textwidth}{!}{\includegraphics{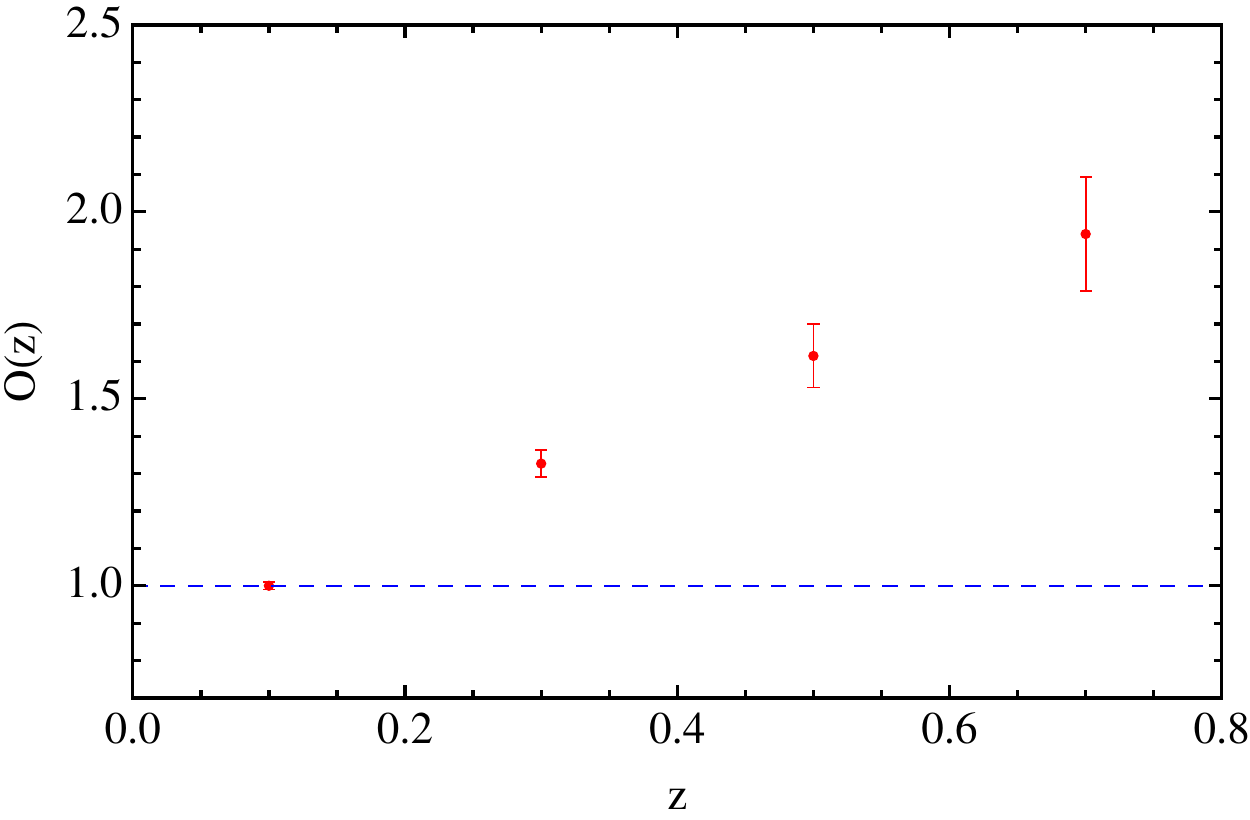}}}\\
\caption{The results for the null-test by binning the mock catalogs for the $H(z)$ and \fs8 data. Left panel: the $w$CDM with dark energy perturbations (cyan circle), the $f(R)$  (red square) and the $f(G)$ (purple diamond).  Right panel: the LTB model.
\label{fig:mocks1}}
\end{figure*}

In what follows we report the results obtained by binning the data in the mock catalogs that we created using different cosmologies. We use the null-test valid for the \lcdm model given by Eq.~(\ref{nulltest2}) and we analyze the mocks. In practice we ask ourselves the following: if the Universe is different from $\Lambda$CDM, how accurately we can test it? As we are analyzing mocks created using a cosmology different from the \lcdm we expect the null-test to fail, i.e. to be different from unity at all the redshifts.

To analyze the mock data we decided to use two different binning: first, we used  $4$ bins from redshift $0$ until redshift $0.8$ to compare them with the results from the actual data; second, we used $10$ bins using all the data, i.e. we extended our analysis up to $z=2.0$.
As both catalogs contain the same number of points and they are uniformly distributed, the mean redshift in each bin will be the same for each cosmology. In Tab.~\ref{tab:Nulltest-results-binning-mocks} we report the values of the $\mathcal{O}(z)$ for the different cosmologies in the $4$-bins case. In the same table, next to each value of the null-test, we present the confidence level, i.e. how many sigmas the values of the null-test are from unity, if the value is smaller than $1$ then the value of the null-test $\mathcal{O}(z)$ is within $1\sigma$ close to unity, if the value is large than one, then the value corresponds to the sigmas away the null-test is.

In Fig.~\ref{fig:mocks1} we show the result for four cosmologies\footnote{We excluded \lcdm for sake of space} and in Tab.~\ref{tab:Nulltest-results-binning-mocks} we report the values found for the null-test, the corresponding errors and the confidence level.

If we test the \lcdm mock catalogs, we get a result that it consistent with $1$ already at $1\sigma$, see Tab.~\ref{tab:Nulltest-results-binning-mocks}, when we use a different mock catalog for instance the $w$CDM one then $\mathcal{O}(z)$ is less than $1$ at more than $2\sigma$'s at almost any redshift, which is the result that we would expect as the growth of the matter density contrast increases because of the dark energy perturbations.
Using the $f(R)$ mocks the null-test gives values closer to unity indicating that it will be more difficult to differentiate the \lcdm and the $f(R)$ model; this is due to the fact that the $f(R)$ model used in this paper has a Hubble parameter which is exactly \lcdm and an $\alpha$ of $0.002$, hence the modification to the growth $f\sigma_8(z)$ is small.
When we use the mocks from $f(G)$ and LTB cosmologies, which both models give substantially different behavior of the Hubble parameter and the growth of matter density contrast, the deviation from unity of the null-test $\mathcal{O}(z)$ becomes more evident, in fact we found that the $f(G)$ can be ruled out at more than $4\sigma$'s  and the LTB at more than $9\sigma$'s.
The results up to $z=2$ can be found in Tab.~\ref{tab:Nulltest-results-binning-mocks-10bins}.

\subsection{Model testing}

\begin{figure*}
\centering
\vspace{0cm}\rotatebox{0}{\vspace{0cm}\hspace{0cm}\resizebox{0.45\textwidth}{!}{\includegraphics{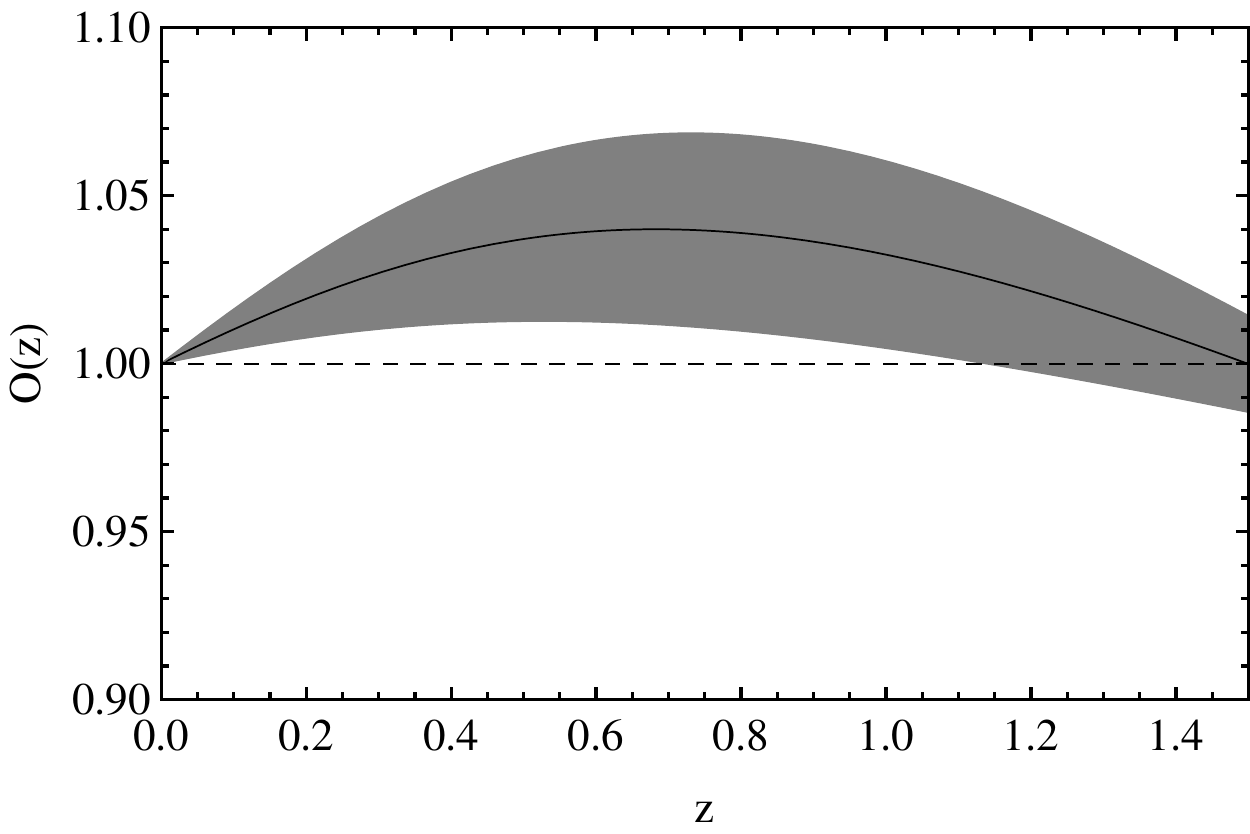}}}
\vspace{0cm}\rotatebox{0}{\vspace{0cm}\hspace{0cm}\resizebox{0.45\textwidth}{!}{\includegraphics{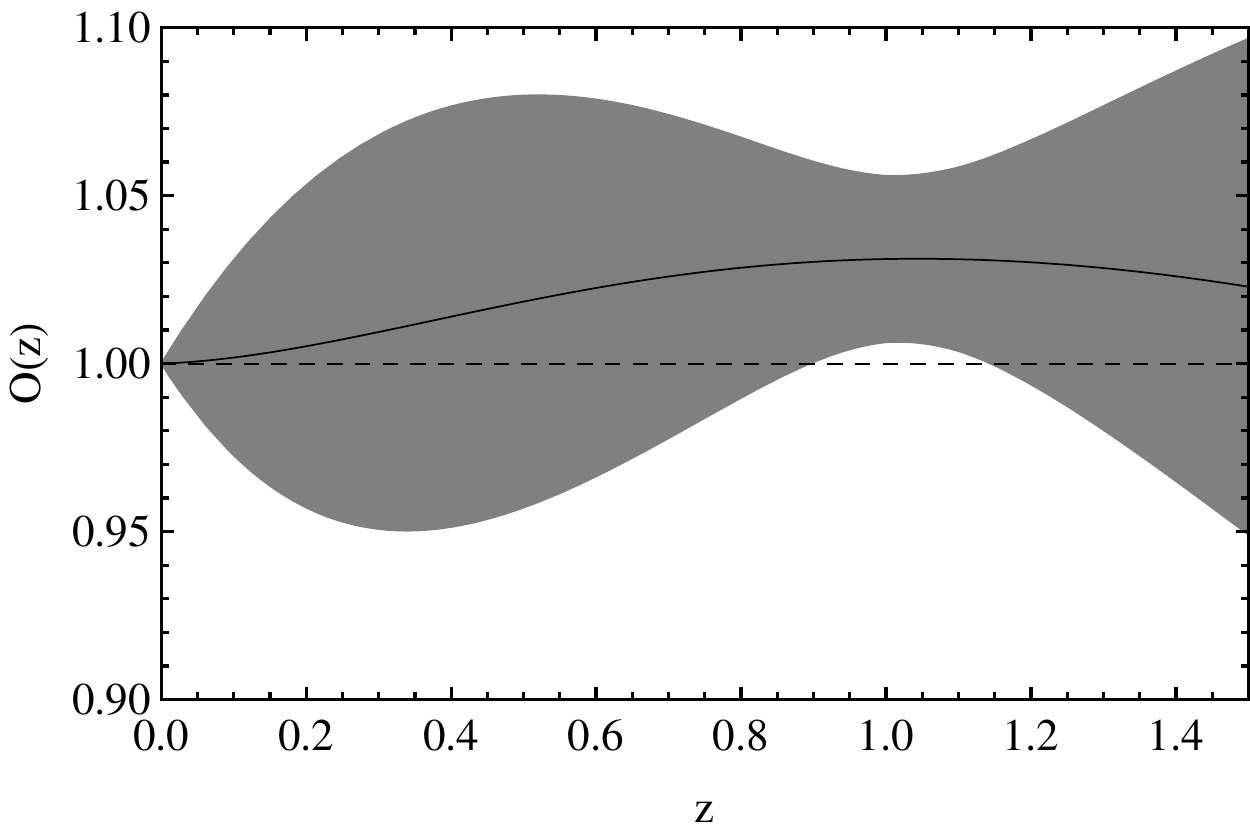}}}
\caption{The results for the null-test for the \lcdm (left) and $w$CDM (right) using actual data for the $H(z)$ and \fs8 data.
\label{fig:modelfittingreal}}
\end{figure*}

An interesting alternative to binning is to fit the data, either real or mock, to the $\Lambda$CDM and $w$CDM models and then reconstruct the null-test $\mathcal{O}(z)$. In Fig. \ref{fig:modelfittingreal} we show the results of reconstructing the null-test with the real data fitted by the \lcdm (left) and $w$CDM (right) models respectively. Clearly, the null-test as reconstructed with the real data seems to be compatible with unity at the $1.5\sigma$ level.

Next, we will also test how well the null-test will be reconstructed with future data. For this, we also consider the different cosmologies mentioned in a previous section and fit the mocks with both the \lcdm and $w$CDM models. The reason for this is that we want to make a direct test of the standard cosmological model with as few extra assumptions as possible. We should stress that in this case any deviation from unity implies a breakdown of either the fundamental assumptions of the standard cosmological model, i.e. homogeneity, the validity of GR etc, or that the DE models used are not a good description of the data.

\begin{figure*}
\centering
\vspace{0cm}\rotatebox{0}{\vspace{0cm}\hspace{0cm}\resizebox{0.45\textwidth}{!}{\includegraphics{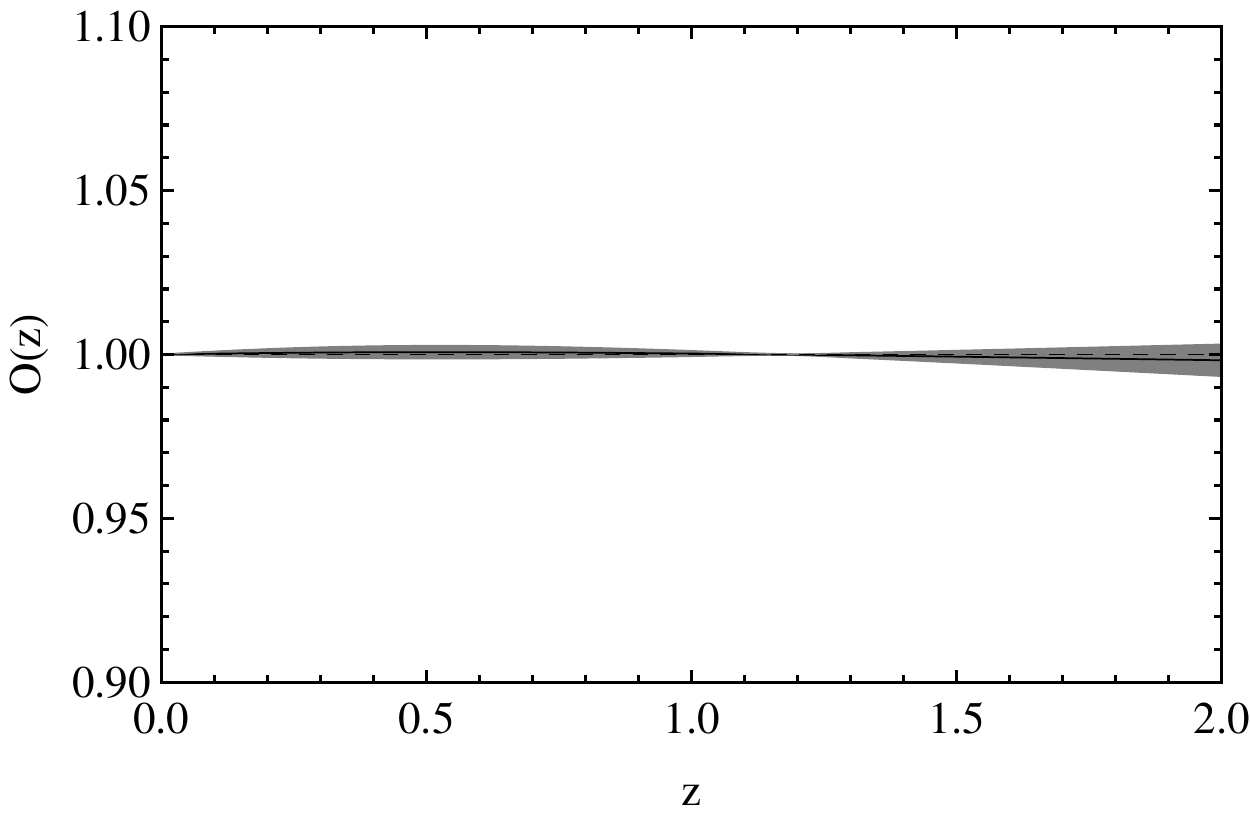}}}
\vspace{0cm}\rotatebox{0}{\vspace{0cm}\hspace{0cm}\resizebox{0.45\textwidth}{!}{\includegraphics{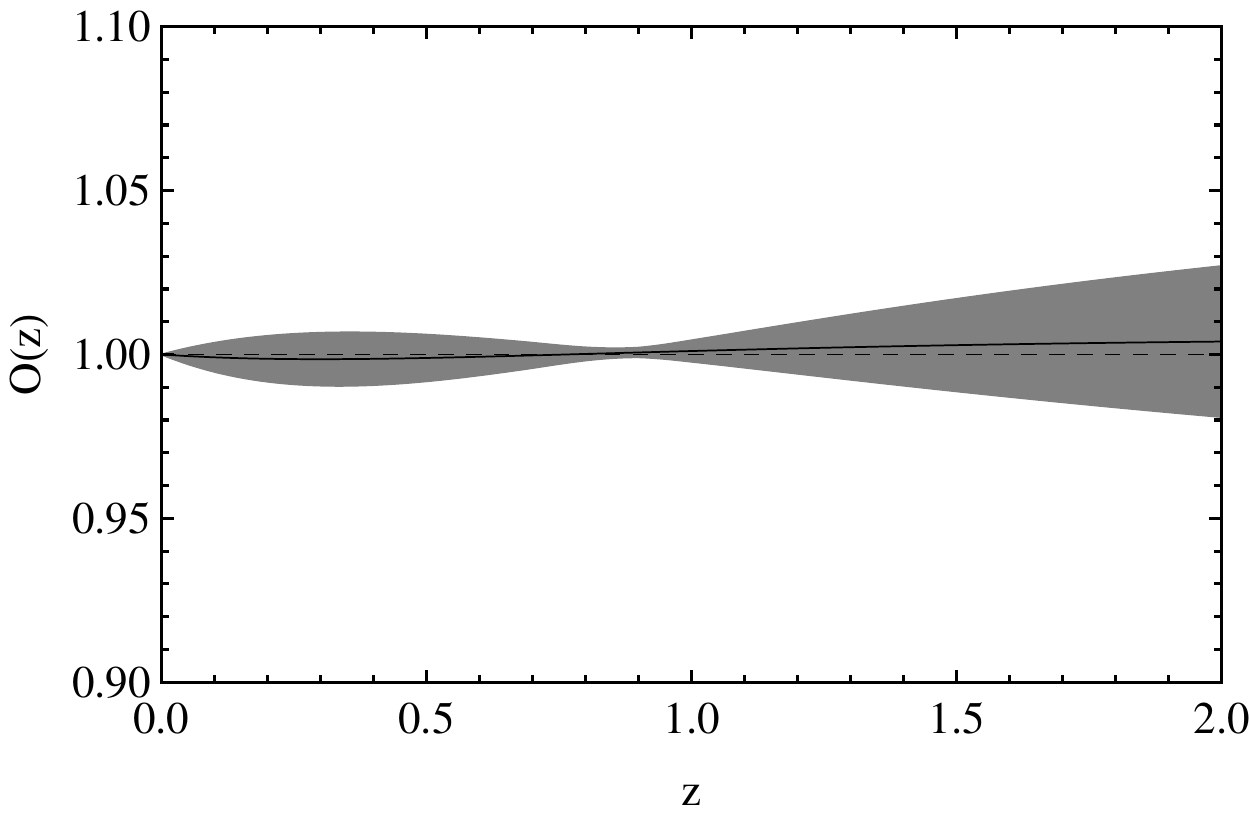}}}\\
\vspace{0cm}\rotatebox{0}{\vspace{0cm}\hspace{0cm}\resizebox{0.45\textwidth}{!}{\includegraphics{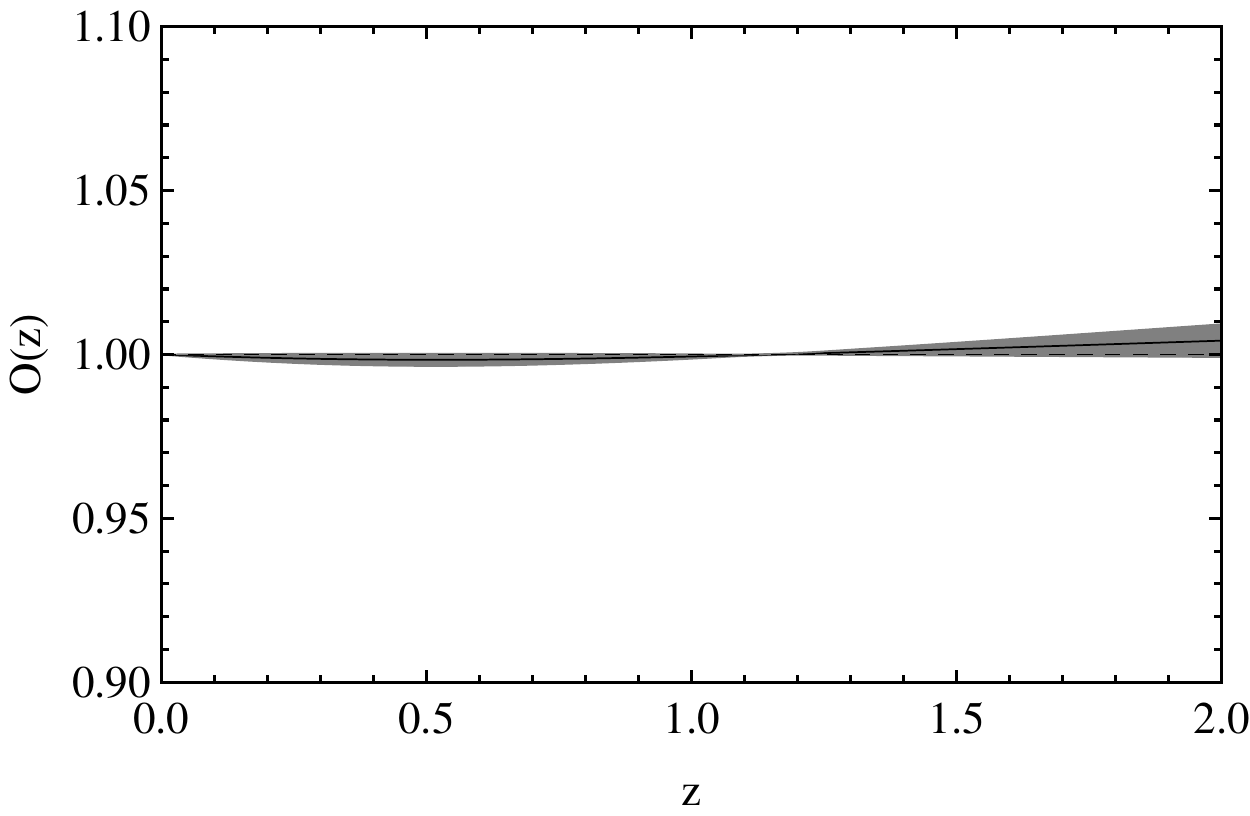}}}
\vspace{0cm}\rotatebox{0}{\vspace{0cm}\hspace{0cm}\resizebox{0.45\textwidth}{!}{\includegraphics{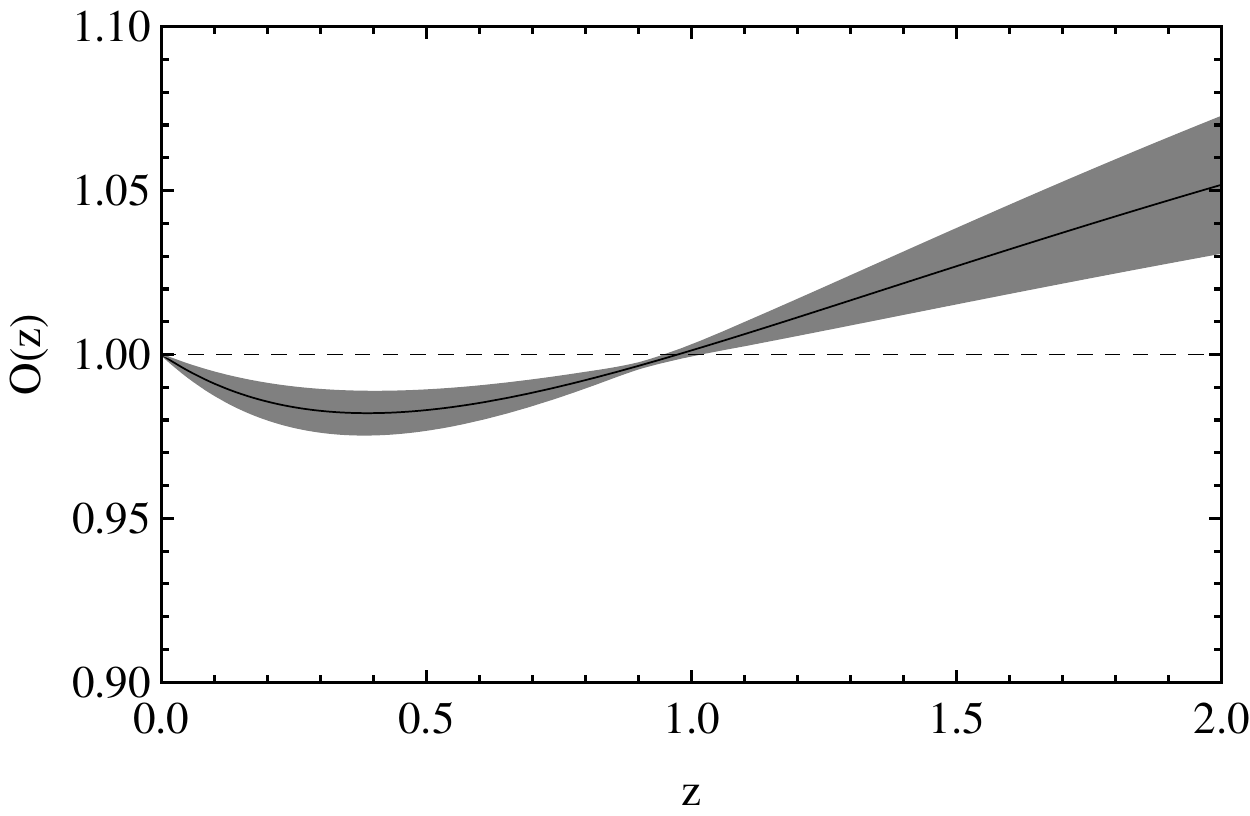}}}\\
\caption{The results for the null-test for the \lcdm (left) and $w$CDM (right) using \lcdm mocks (first row) and the DE perturbations (second row) for the $H(z)$ and \fs8 data.
\label{fig:modelsmocks1}}
\end{figure*}

\begin{figure*}
\centering
\vspace{0cm}\rotatebox{0}{\vspace{0cm}\hspace{0cm}\resizebox{0.45\textwidth}{!}{\includegraphics{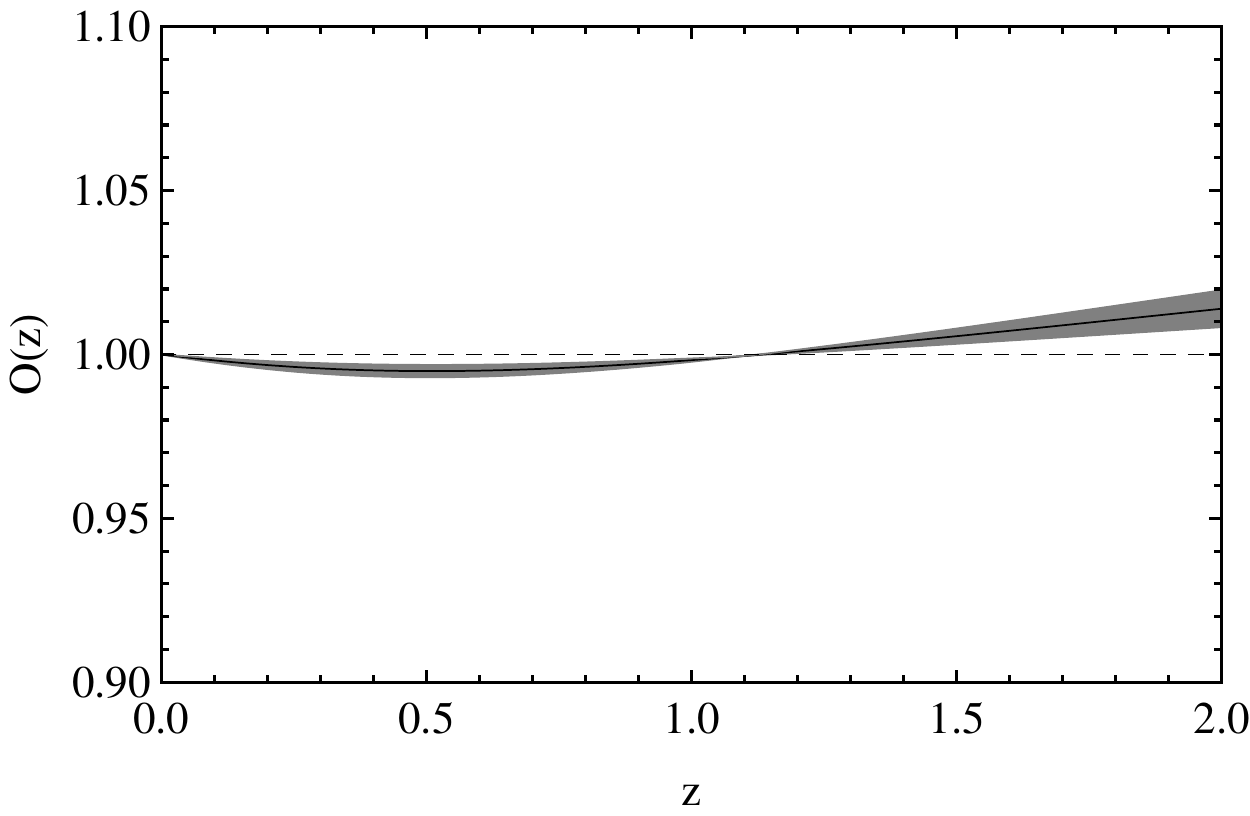}}}
\vspace{0cm}\rotatebox{0}{\vspace{0cm}\hspace{0cm}\resizebox{0.45\textwidth}{!}{\includegraphics{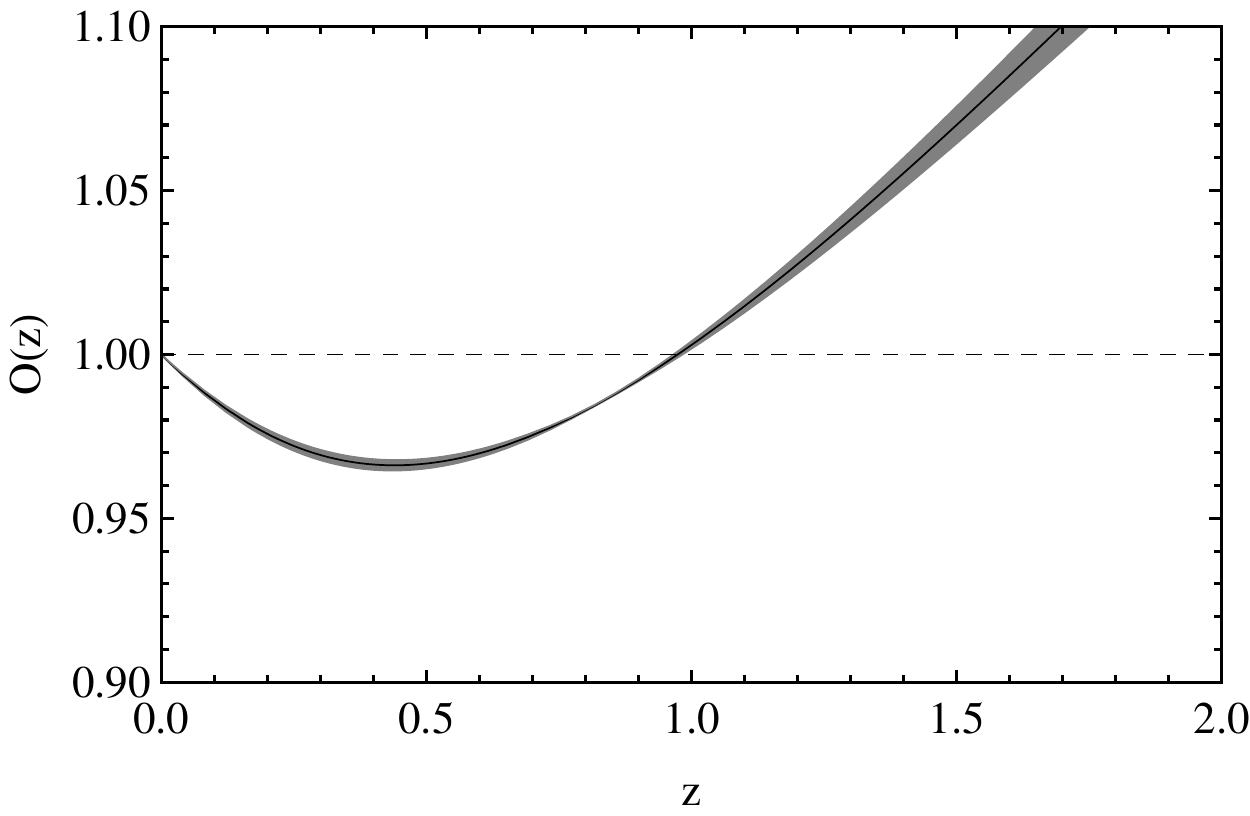}}}\\
\vspace{0cm}\rotatebox{0}{\vspace{0cm}\hspace{0cm}\resizebox{0.45\textwidth}{!}{\includegraphics{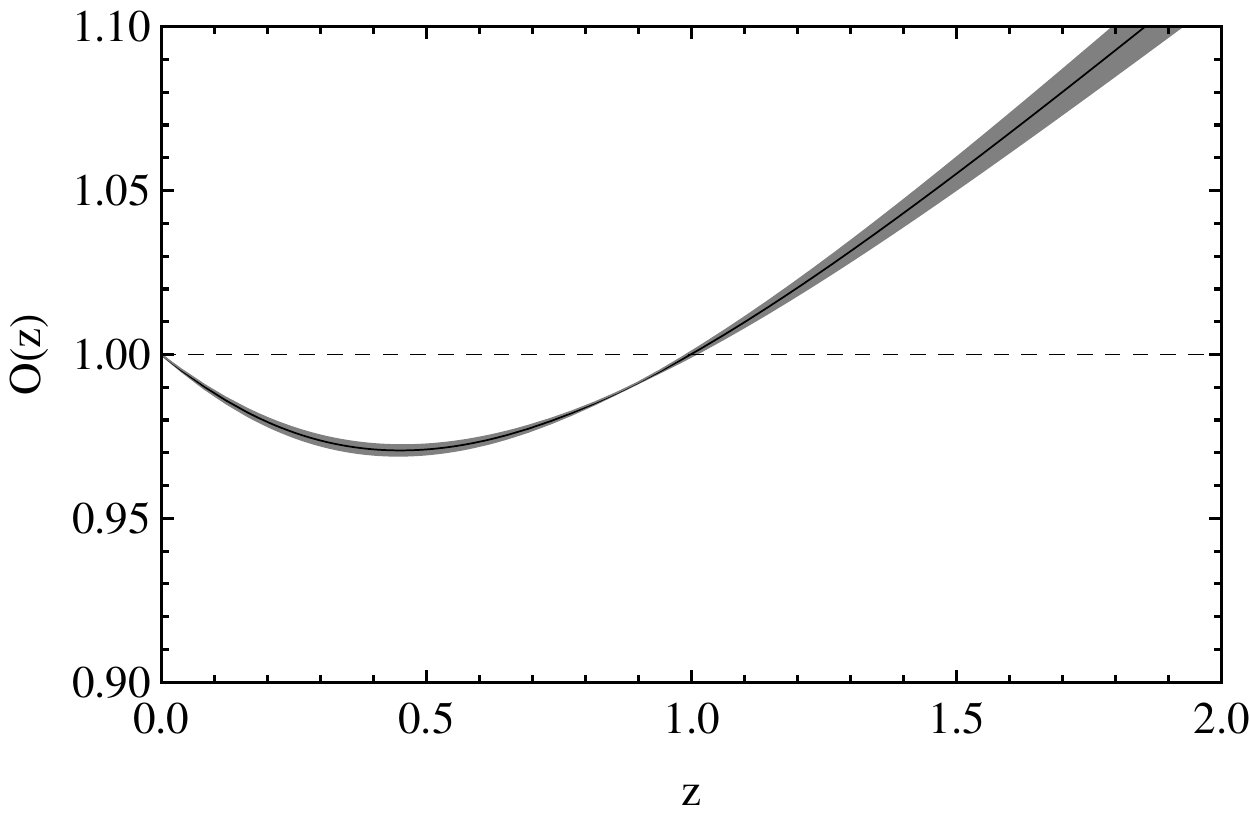}}}
\vspace{0cm}\rotatebox{0}{\vspace{0cm}\hspace{0cm}\resizebox{0.45\textwidth}{!}{\includegraphics{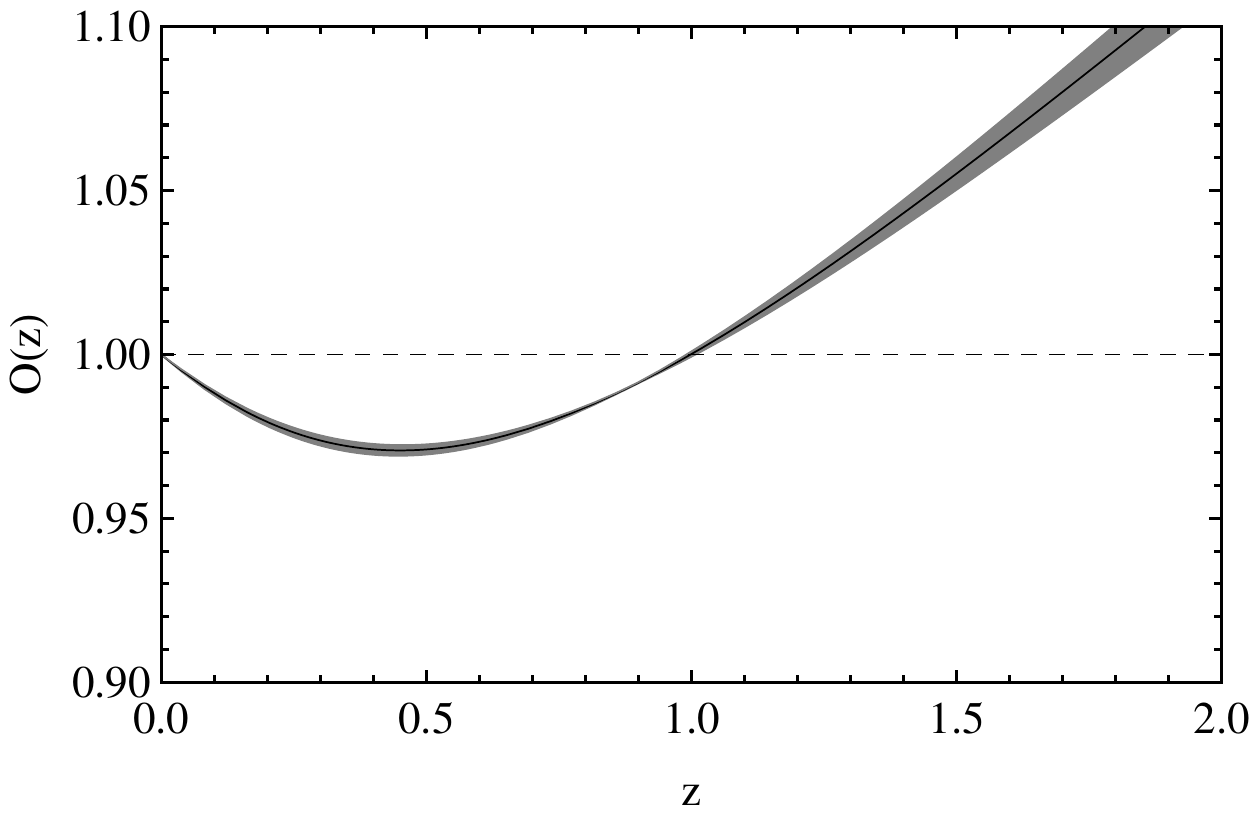}}}\\
\caption{The results for the null-test for the \lcdm model for the mock $H(z)$ and \fs8 data based on the $f(R)$ model for $\alpha=0.002$ (first row left) and $\alpha=0.2$ (first row right). On the second row we show the results for the $f(G)$ $H(z)$ and \fs8 data for the  \lcdm model (left) and  $w$CDM (right)  models respectively.
\label{fig:modelsmocks2}}
\vspace{0cm}\rotatebox{0}{\vspace{0cm}\hspace{0cm}\resizebox{0.45\textwidth}{!}{\includegraphics{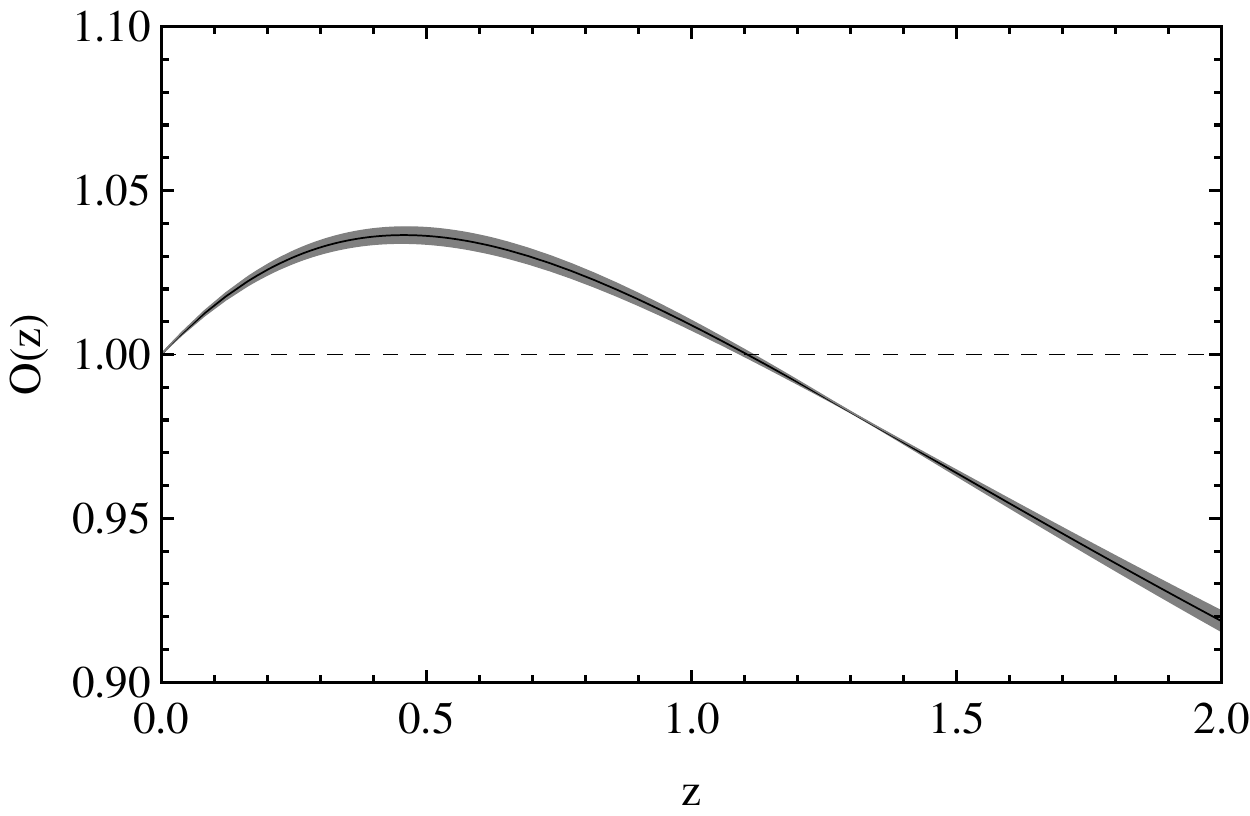}}}
\vspace{0cm}\rotatebox{0}{\vspace{0cm}\hspace{0cm}\resizebox{0.45\textwidth}{!}{\includegraphics{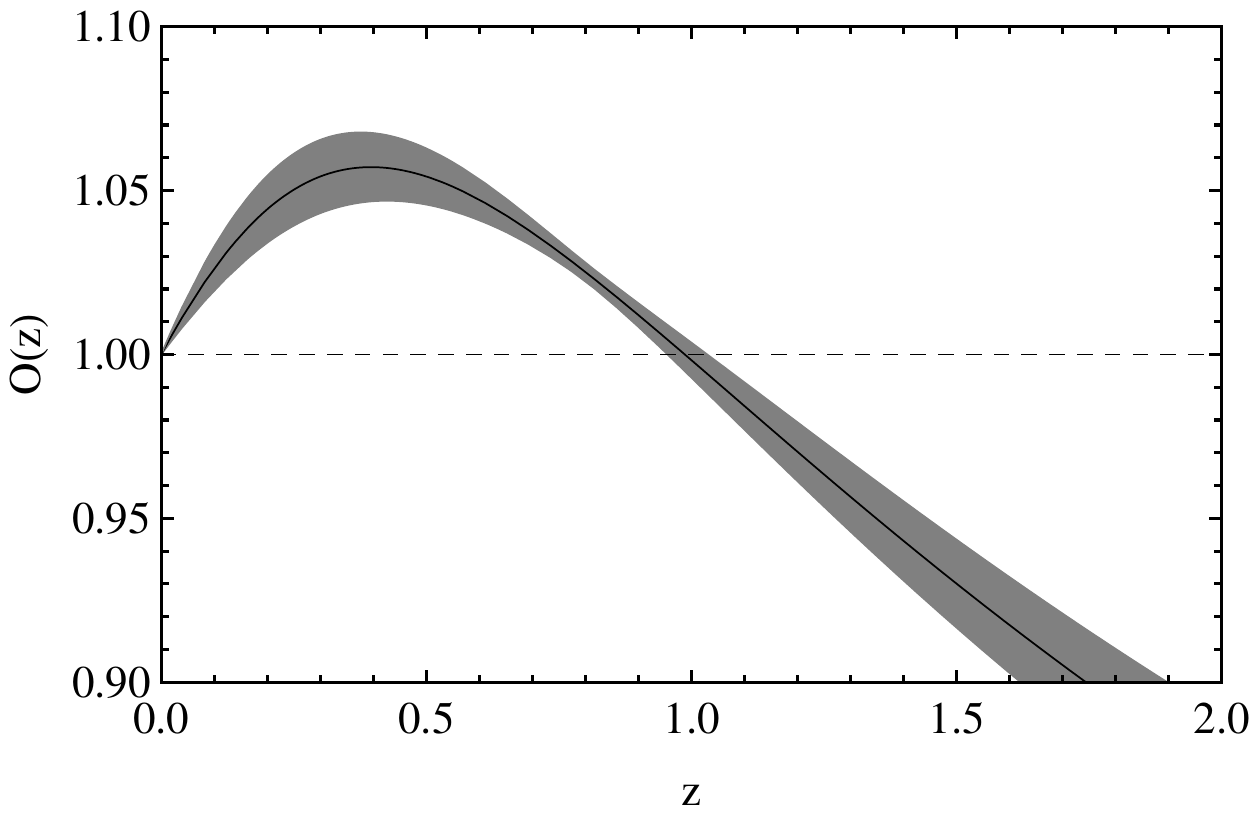}}}
\caption{The results for the null-test for the LTB $H(z)$ and \fs8 mocks fitted with the \lcdm (left) and  $w$CDM (right). \label{fig:LTBmocks}}
\end{figure*}

In Fig. \ref{fig:modelsmocks1} we show the results for the null-test for the \lcdm (left) and $w$CDM (right) using \lcdm mocks (first row) and the DE perturbations (second row) for the $H(z)$ and \fs8 data. In Fig. \ref{fig:modelsmocks2} we show the results for the null-test for the \lcdm model for the mock $H(z)$ and \fs8 data based on the $f(R)$ model for $\alpha=0.002$ (first row left) and $\alpha=0.2$ (first row right). On the second row we show the results for the $f(G)$ $H(z)$ and \fs8 data for the  \lcdm model (left) and  $w$CDM (right)  models respectively. Finally, in Fig. \ref{fig:LTBmocks} we show the results for the null-test for the LTB $H(z)$ and \fs8 mocks fitted with the \lcdm (left) and  $w$CDM (right).

We find that the $\mathcal{O}(z)$ null-test will be particularly successful at detecting deviations from GR at high significance $(\gtrsim5\sigma)$, especially of the $f(R)$ and $f(G)$ types (Fig.~\ref{fig:modelsmocks2}), but also deviations from the FRW metric (Fig.~\ref{fig:LTBmocks}). This is due to the fact that these models have significantly different evolution for the matter density perturbations, which is encoded in the $\Geff$ and can be detected by the null-test.

\begin{figure}[t]
\centering
\vspace{0cm}\rotatebox{0}{\vspace{0cm}\hspace{0cm}\resizebox{0.45\textwidth}{!}{\includegraphics{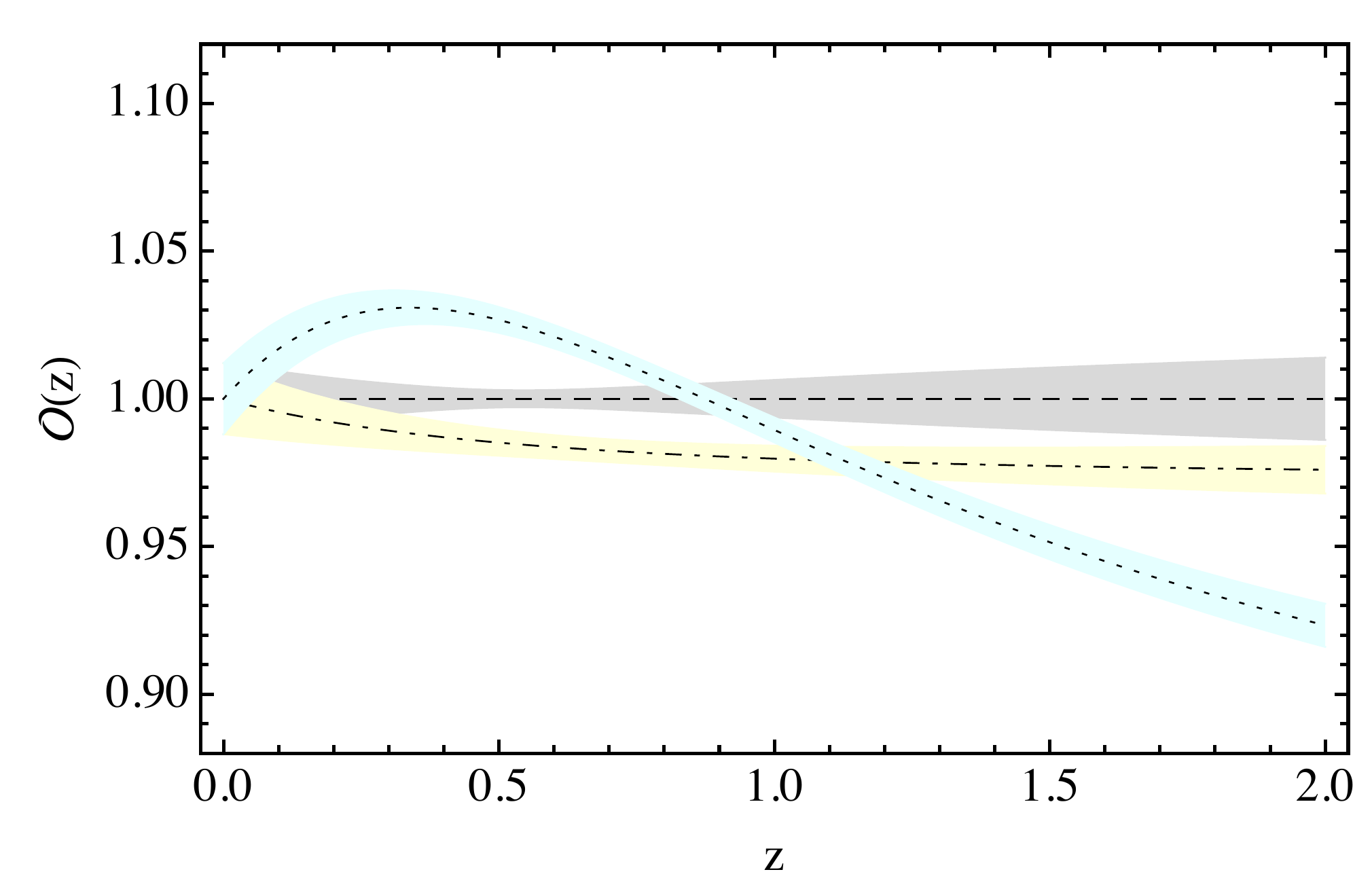}}}
\caption{The $\mathcal{O}(z)$ for three different cosmologies: black dashed line for $\Lambda$CDM mock catalogs (light gray area is the error), black dot-dashed line for $w$CDM mock catalogs (light yellow area is the error), black dotted line for the mixing cosmology, i.e. using $H(z)$ mock under for $w$CDM and $f\sigma_8(z)$ mock for  $\Lambda$CDM (light cyan area is the error).
\label{fig:tension-in-data}}
\end{figure}

\subsubsection{Alternative data and theories}

As an extra check we also use alternative data instead of just the $H(z)$, namely the Supernovae type Ia (SnIa) to reconstruct the Hubble parameter. In particular we used the latest {\em Union 2.1}\footnote{The SnIa data can be found in http://supernova.lbl.gov/Union/ and in \cite{Suzuki:2011hu}} set of 580 SnIa data of Suzuki et al. \cite{Suzuki:2011hu} that spans  from redshift $0.015$ up to $1.4$.

\begin{figure*}
\centering
\vspace{0cm}\rotatebox{0}{\vspace{0cm}\hspace{0cm}\resizebox{0.45\textwidth}{!}{\includegraphics{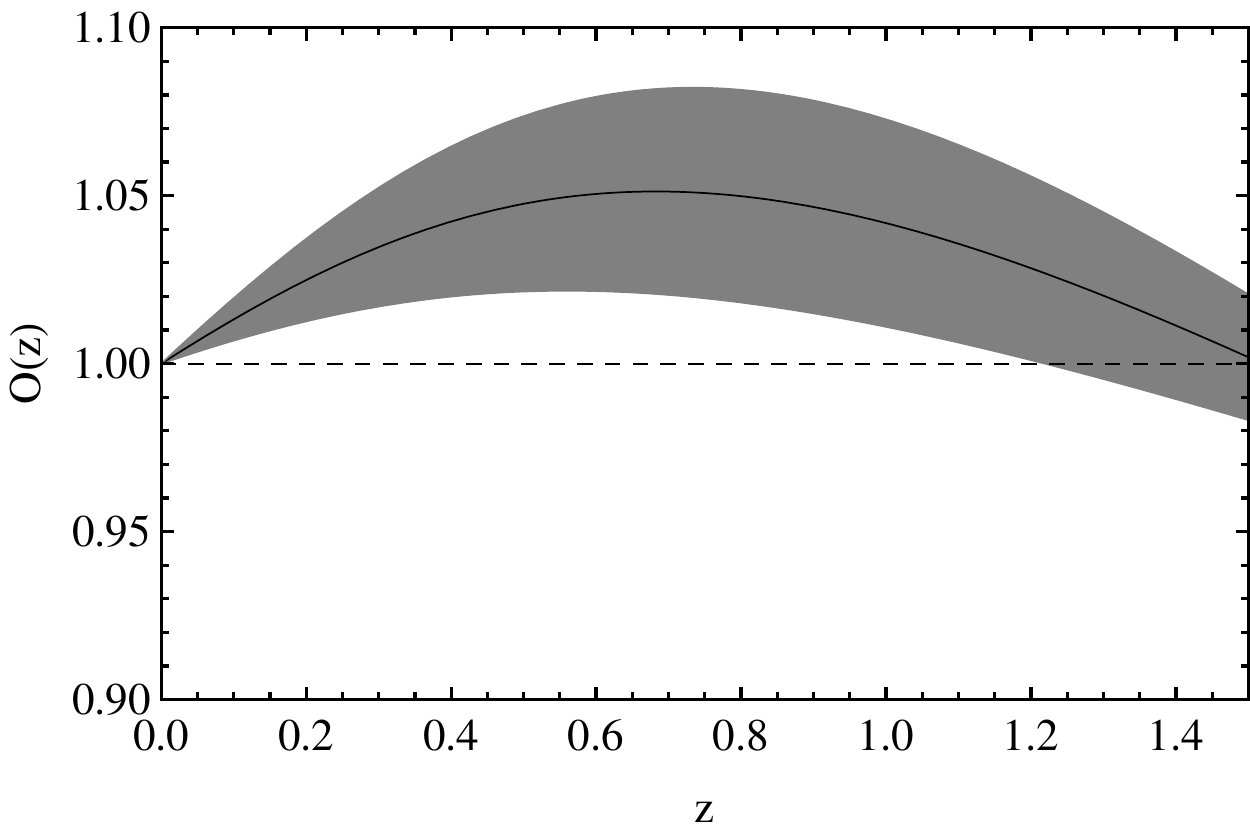}}}
\vspace{0cm}\rotatebox{0}{\vspace{0cm}\hspace{0cm}\resizebox{0.45\textwidth}{!}{\includegraphics{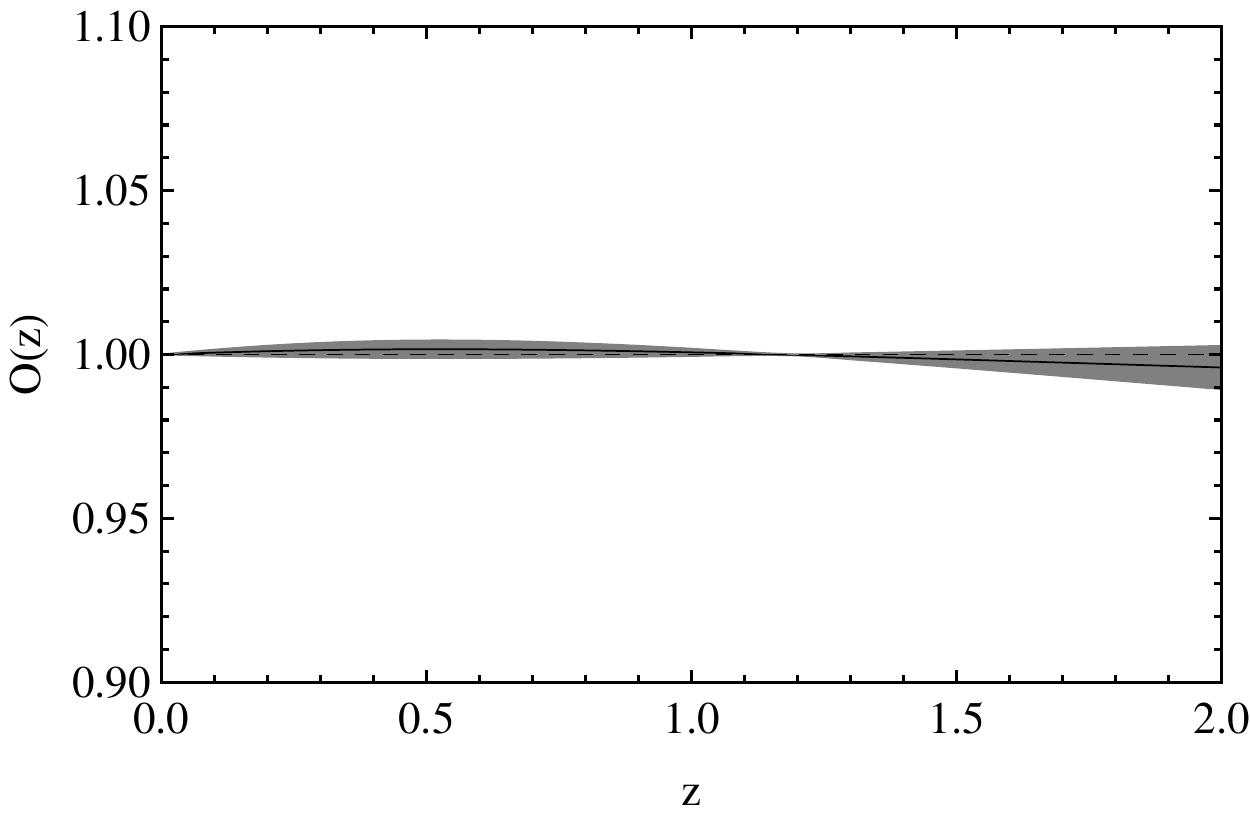}}}\\
\caption{The results for the null-test for the \lcdm for the real data (left) and the mock data (right) for the SnIa and \fs8 data.
\label{fig:LCDMSnIarealvsmock}}
\end{figure*}

\begin{figure*}
\centering
\vspace{0cm}\rotatebox{0}{\vspace{0cm}\hspace{0cm}\resizebox{0.45\textwidth}{!}{\includegraphics{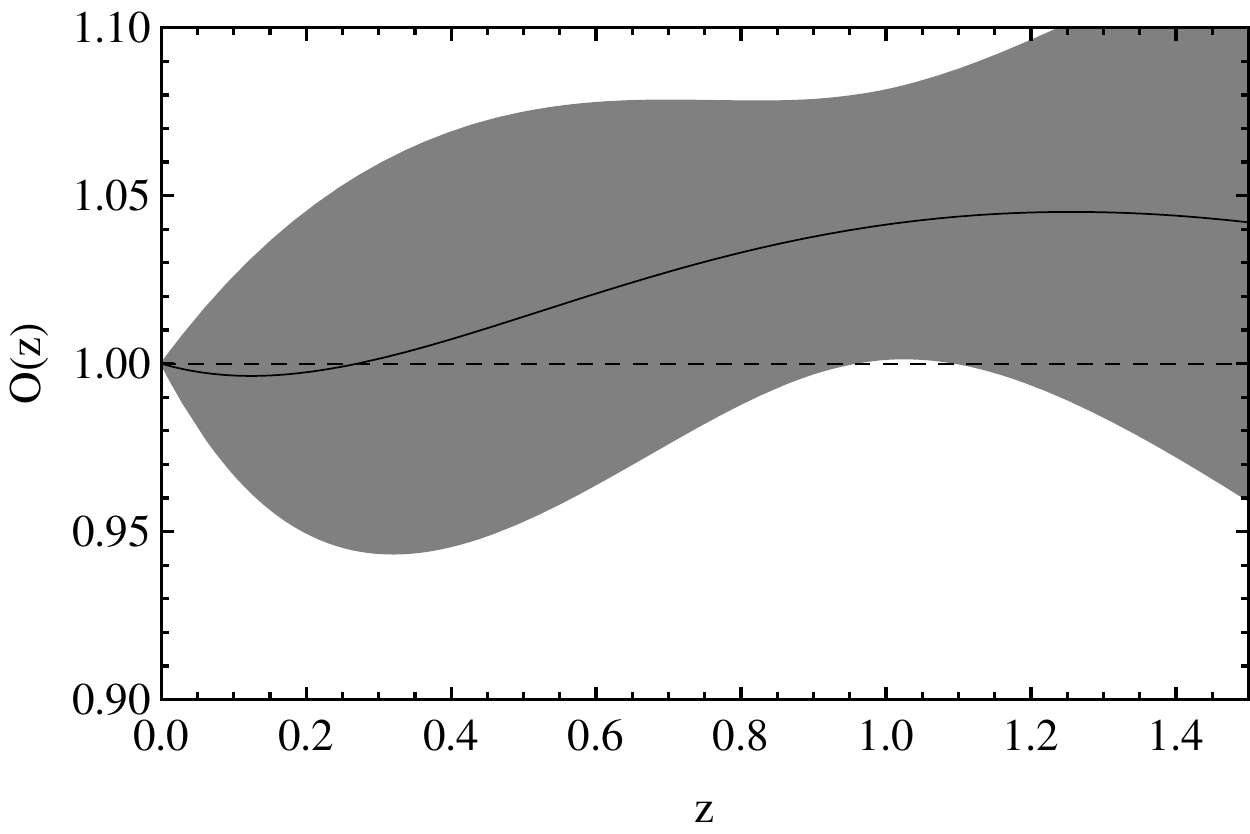}}}
\vspace{0cm}\rotatebox{0}{\vspace{0cm}\hspace{0cm}\resizebox{0.45\textwidth}{!}{\includegraphics{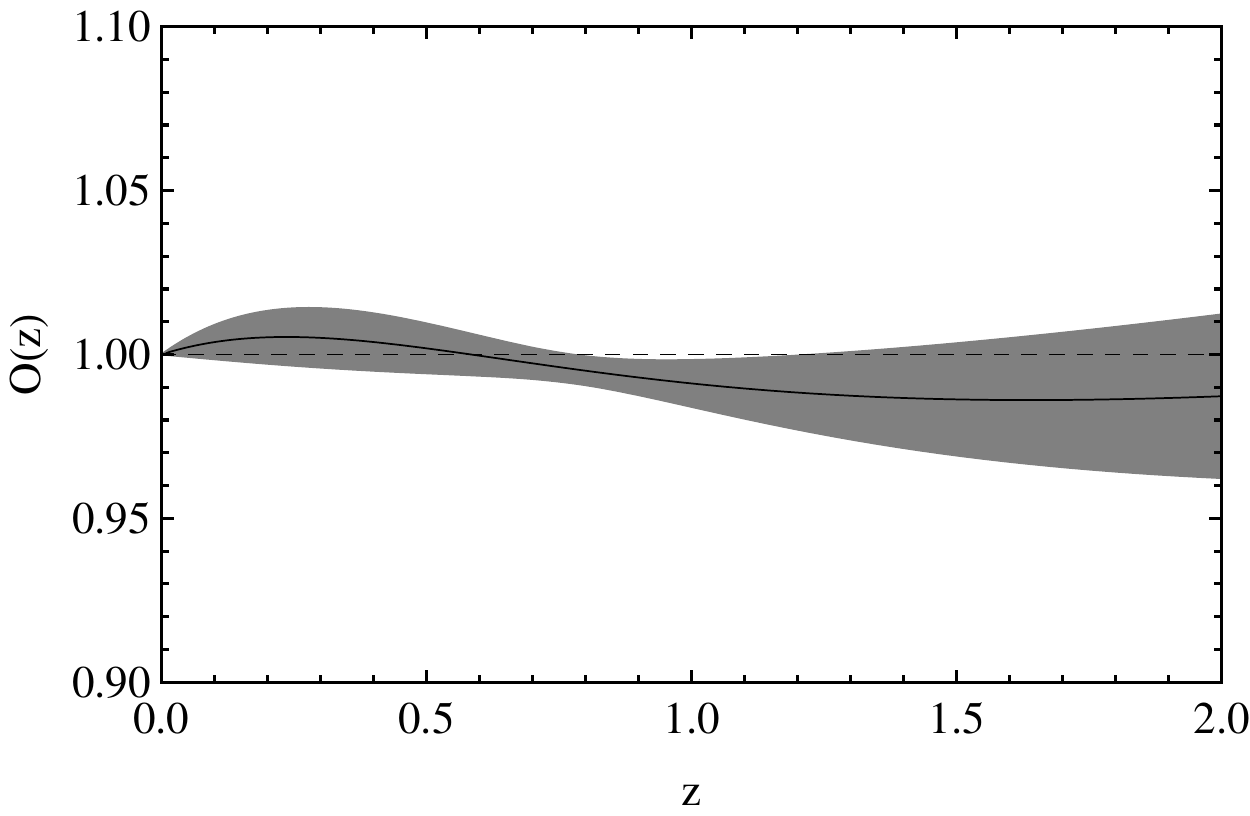}}}\\
\caption{The results for the null-test for the $w$CDM for the real data (left) and the mock data (right) for the SnIa and \fs8 data.
\label{fig:wCDMSnIarealvsmock}}
\end{figure*}

The results for this reconstruction are shown in Fig.~\ref{fig:LCDMSnIarealvsmock} for the \lcdm and in Fig.~\ref{fig:wCDMSnIarealvsmock} for the $w$CDM. We find that they are in excellent agreement with that of the $H(z)$ data shown earlier, thus eliminating any possibility of bias due to the use of the particular data used to reconstruct the Hubble expansion history.

\begin{figure*}
\centering
\vspace{0cm}\rotatebox{0}{\vspace{0cm}\hspace{0cm}\resizebox{0.45\textwidth}{!}{\includegraphics{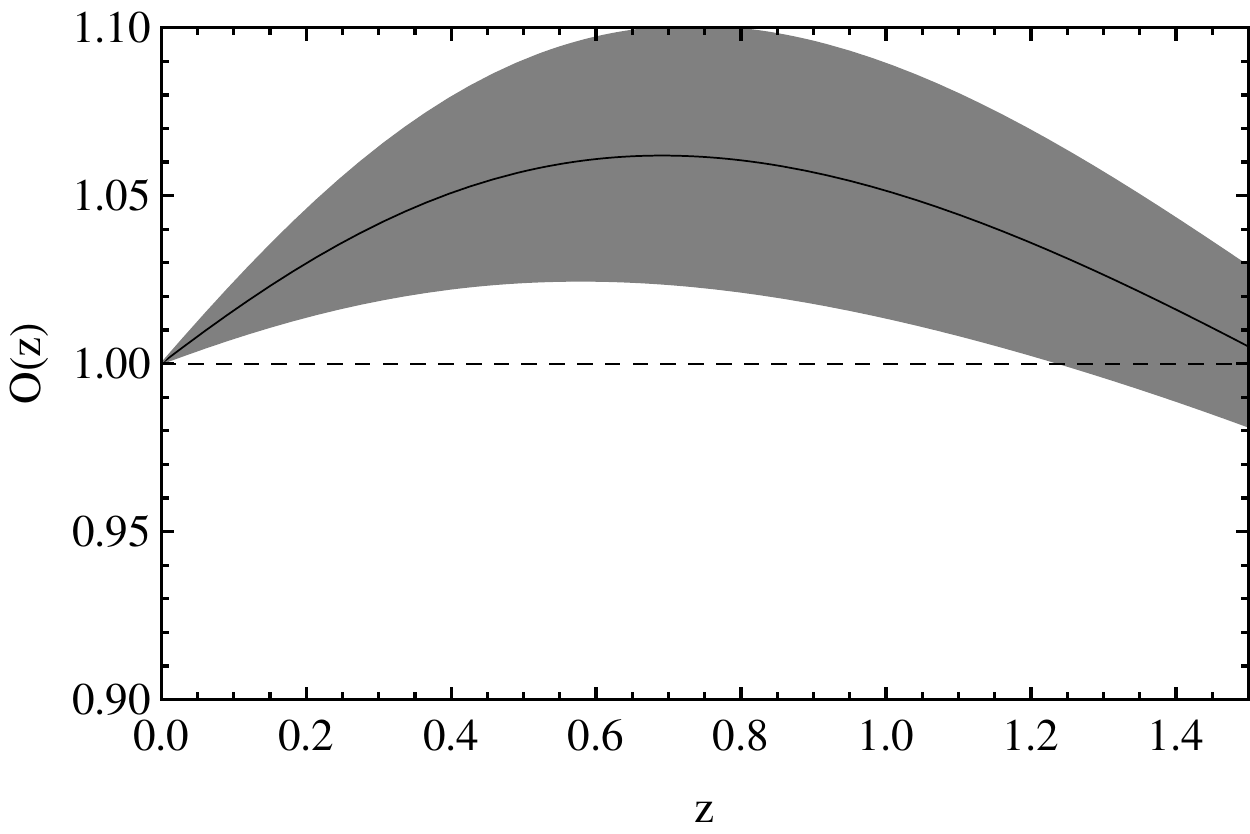}}}
\vspace{0cm}\rotatebox{0}{\vspace{0cm}\hspace{0cm}\resizebox{0.45\textwidth}{!}{\includegraphics{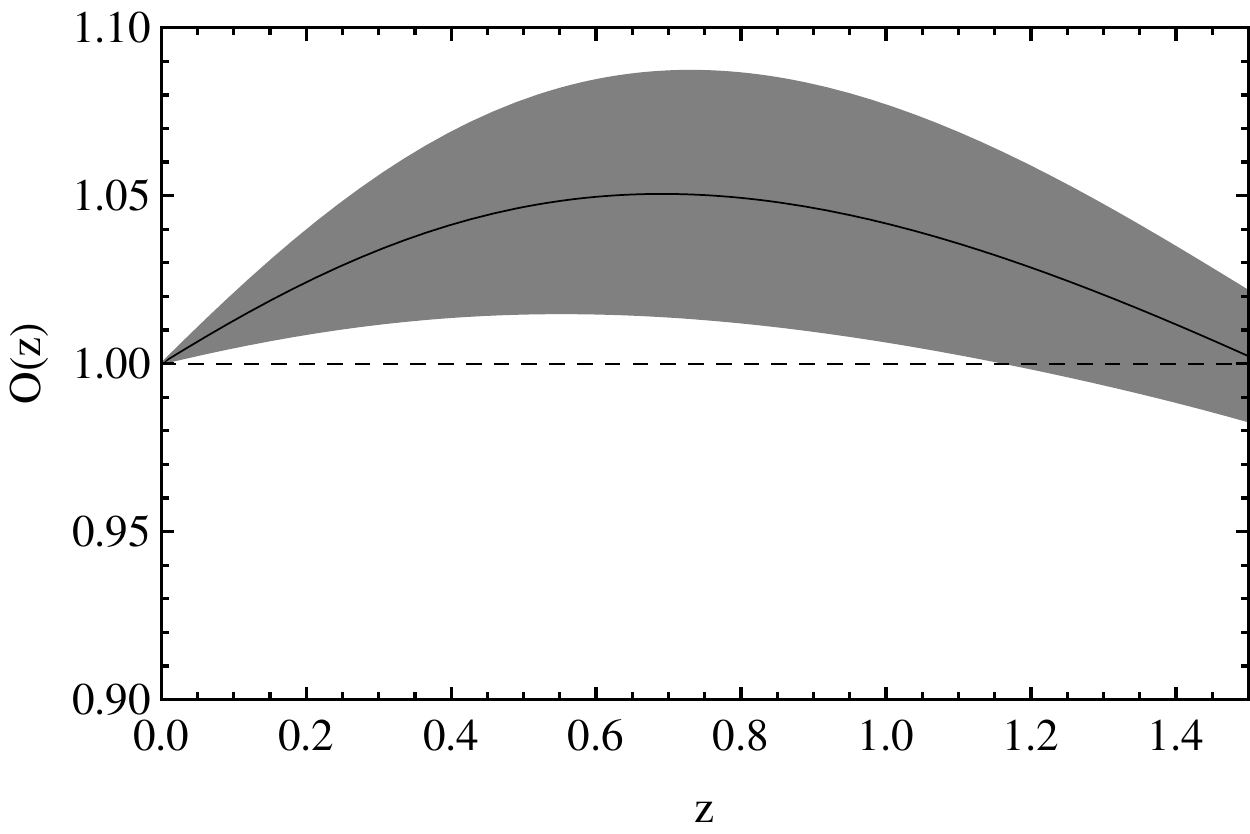}}}
\caption{The results for the null-test for the $f(R)$ model for the real SnIa and \fs8 data (left) and the $H(z)$ and \fs8 data data (right).
\label{fig:fRrealSnIavsHz}}
\end{figure*}

Finally, we also consider $f(R)$ models in the reconstruction of the null-test. Specifically, in Fig. \ref{fig:fRrealSnIavsHz} we show the results for the null-test for the $f(R)$ model for the real SnIa and \fs8 data (left) and the $H(z)$ and \fs8 data data (right). Again the results are in good agreement, thus demonstrating that the null-test is not particularly sensitive on the model used.

\subsection{Validity of the growth-rate data used with the null-test}

It should be noted that unfortunately, none of the $f\sigma_8(z)$ data used in this analysis has been evaluated in a completely model independent manner. Usually an underlying cosmology has to be considered in order to extract information about the growth rate parameter, and most of the time this cosmology is the $\Lambda$CDM. Another limitation of the $f\sigma_8$ data is that part of the measurement comes from a range of wave numbers that falls into the nonlinear regime, with the typical scales ranging from $30$ up to $200$ Mpc/h, which expressed as a wavenumber is $k = 0.03 - 0.2$ h/Mpc. The maximum wavenumber that we can reach before entering into the non linear regime depends on the redshift of the measurements; starting from $k_{max}= 0.1$ h/Mpc at $z=0$ up to $k_{max}=0.2$ h/Mpc at high redshifts (above $1$).

Most surveys \cite{hudson, samushia} already do take into account non-linearities via the non linear BAO diffusion. The latter is usually treated as a ``nuisance'' parameter, which will be marginalized over as it does not depend on the cosmology, and by doing so the cosmological information comes only from the linear part of the power spectrum. Moreover, the growth rate $f(z)=d\ln\delta/d\ln a$ is a linear quantity because is directly proportional to the velocity perturbations $\delta'$ with no scale dependence. Clearly there is still a lot room for improvement in order to have completely model independent and better measurements of the growth index. It should be stressed however, that this problem with the data could potentially be resolved in the near future if the data can be created with more model independent techniques, thus alleviating this constraint from our null-test.

\subsection{Tension in the data}
In this subsection we will briefly discuss another issue, that of the potential problem that the $H(z)$ data might originate from a different cosmology but the growth $f\sigma 8$ data has been evaluated under the assumption of \lcdmb. For instance, the later time scale independent growth could be generated by $f(R)$ gravity theory, but some of analysis assumed the coherent growth and then what it is measured is the mean $f\sigma 8(z)$ in the valid $k$ bins. Above all, uncertainties due to the non-linear or non-perturbed terms prevail through those measurements and as a result, we are not able to test anything other than simple GR models. Therefore, if one wants to constrain a specific theoretical model, such as $f(R)$ or $f(G)$, he/she should analyze the data based upon the model-dependent method.

However, our test is very sensitive to variations of cosmologies or, in other terms, in tensions in the data, as in this particular case.  An interesting game-test is the following: Let us imagine we have a data set of $H(z)$ measured in a model independent way and they favor a $w$CDM model; then we have a different data set of $f\sigma8(z)$ measured assuming a model (in this case \lcdmb) so these data are clearly biased as they will favor a \lcdm model rather than the true underlying cosmology (which is $w$CDM with perturbations also in the dark energy sector). So we can think to use these two different data sets with the null-test: in practice, in order to quantify the sensitivity of our null-test, we use a mock catalog for H(z) using the $w$CDM model and another mock catalog using for $f\sigma8(z)$ using the \lcdm model. We notice that the null-test deviates from unity at the level of about $10\%$ ($\mathcal{O} \sim 1.1$) and in average of about $4.0\%$.

We can also ask how much $f\sigma8(z)$ itself deviates using the wrong cosmology and test which is more sensitive. We find that for $f\sigma8(z)$ for the two different cosmologies, i.e. for \lcdm and $w$CDM, the difference between the $f\sigma8(z)$ evaluated using two different cosmologies is of about $2-5\%$ and in average of about $2.4\%$.

We can also have a closer look at the errors on the null-test mixing-up two different cosmologies. Implementing the full error propagation analysis we realize that the errors remain basically unchanged when we mix up the two different cosmologies; the reason is that the error propagation formula does not change and the important quantities are the errors on the measured quantity, i.e. $H(z)$, $f\sigma_8(z)$, $\omega_m$ and $\sigma_8$. In Fig.~\ref{fig:tension-in-data} we plot the null-test using $\Lambda$CDM and $w$CDM mock catalogs and the null-test mixing the cosmologies.

\section{Conclusions}\label{sec:conclusions}
In this paper we have reconstructed the null-test, that can be used to probe potential deviations from \lcdm and was first presented in Ref.~\cite{Nesseris:2014mfa}, by using the $H(z)$, SnIa and \fs8 data. We performed the reconstruction in two different ways: by directly binning the data and by fitting the data to various dark energy models like the \lcdm and $w$CDM and then calculating the null-test. We find that both methods have different advantages; the former uses as few assumptions as possible while the latter directly tests the standard cosmological model.

We have also generalized the null-test and extended it for modified gravity models and models with strong DE perturbations, by taking into account the $\Geff$ term in Eq.~(\ref{ode}). We have explicitly checked that when this term is taken into account, then the null-test is constant as expected for modified gravity models. This allows us to verify that deviations from unity in the original version of the null-test presented in Ref.~\cite{Nesseris:2014mfa} can indeed also be attributed to modifications of gravity.

We have found that deviations from unity could be due to several reasons, either new physics including modifications of gravity and strong dark energy perturbations, or breakdowns of one of the basic assumptions of the standard cosmological model, i.e. deviation from the FLRW metric and homogeneity or finally, a possible tension between $H(z)$ (obtained directly or derived) and the $f\sigma_8$ data. In all cases due to the nature of the null-tests and that they have to be constant at all redshifts, it is enough to a have a statistically significant deviation at one redshift to detect one of the above reasons. A possible limitation at the moment is that the null-test cannot tell us which of the above reasons would be responsible for that deviation though. However, our growth null-test will be extremely useful if joined with other null-tests, like the $\Omega_K(z)$ presented in \cite{Clarkson:2007pz} which is able to test the assumptions of homogeneity and isotropy of the Universe.

We also examined how well the null-test can be reconstructed by future data by creating mock catalogs based on a LSST-like survey and on the \lcdm model, a model with strong DE perturbations, the $f(R)$  and $f(G)$ models, and the large void LTB model that exhibit different evolution of the matter perturbations. This was done so as to examine how well our null-test can be reconstructed using the data from upcoming surveys.

Our results were presented in Figs.~\ref{fig:binrealvsmock-null}-\ref{fig:fRrealSnIavsHz}. We found that when reconstructed with real data the null-test is consistent with unity at the $2\sigma$ level, with both the binning and the model testing methods. However, when we reconstruct it with the mock data based on the specifications of a LSST-like survey and various models that go beyond the \lcdm, ie the $f(R)$, $f(G)$ models and the LTB, we find that the null-test can detect deviations from unity at the $5\sigma$ and also $9\sigma$ level.

Overall, the novelty of our null-test is that it can directly test the fundamental assumptions of the standard cosmological model with as few assumptions as possible. Therefore, it will definitely prove to be an invaluable tool in the near future given the plethora of upcoming surveys that will produce high quality data.

\begin{table*}
\begin{centering}
\begin{tabular}{|c|c|c|c|c|c|c|c|c|c|c|}
\hline
\multicolumn{11}{|c|}{MOCK} \tabularnewline
\hline
 &  \multicolumn{2}{c|}{\lcdm}  &  \multicolumn{2}{c|}{$w$CDM}  &  \multicolumn{2}{c|}{$f(R)$}&  \multicolumn{2}{c|}{$f(G)$}&  \multicolumn{2}{c|}{LTB}\tabularnewline
\hline
z  & \OO$(z) \pm \sigma_{\mathcal{O}(z)}$  & $\sigma$'s& \OO$(z) \pm \sigma_{\mathcal{O}(z)}$  & $\sigma$'s& \OO$(z) \pm \sigma_{\mathcal{O}(z)}$  & $\sigma$'s& \OO$(z) \pm \sigma_{\mathcal{O}(z)}$  & $\sigma$'s& \OO$(z) \pm \sigma_{\mathcal{O}(z)}$  & $\sigma$'s \tabularnewline
\hline
$0.1$& $1.000\pm 0.010$ & $0$ & $1.000\pm 0.010$ & $0$& $1.000\pm 0.010$& $0$ & $1.000\pm 0.009$& $0$ & $1.000 \pm 0.010$ & $0$\tabularnewline
\hline
$0.3$ & $ 0.994\pm 0.013$ & $0.447$ & $0.971 \pm 0.013$ & $2.302$& $0.975\pm 0.013$ & $1.924$ & $0.983\pm 0.012$ &$1.428$ &$1.327\pm 0.035$ & $9.191$\tabularnewline
\hline
$0.5$ & $0.989 \pm 0.022$ &$0.498$ &$0.970 \pm 0.021$ & $1.421$ & $0.978\pm 0.022$ & $0.987$ & $0.958\pm 0.020$ &$2.154$ &$1.612\pm0.085$ &$7.250$ \tabularnewline
\hline
$0.7$ & $0.979 \pm 0.032$ &$0.637$ &$0.962\pm 0.031$ & $1.203$ & $0.964\pm 0.032$ & $1.111$ &$0.885 \pm 0.027$ & $4.205$ & $1.941\pm0.153$& $6.151$ \tabularnewline
\hline
$0.9$ & $0.980 \pm 0.045$ &$0.460$ &$0.951\pm 0.041$ & $1.196$ & $0.950\pm 0.043$ & $1.166$ &$0.850 \pm 0.036$ & $4.203$ & $2.218\pm 0.234$& $5.213$ \tabularnewline
\hline
$1.1$ & $0.961 \pm 0.055$ &$0.707$ &$0.947\pm 0.052$ & $1.032$ & $0.935\pm 0.054$ & $1.205$ &$0.801 \pm 0.043$ & $4.597$ & $2.552\pm 0.338$& $4.596$ \tabularnewline
\hline
$1.3$ & $0.968 \pm 0.068$ &$0.464$ &$0.949\pm 0.062$ & $0.821$ & $0.915\pm 0.064$ & $1.329$ &$0.784 \pm 0.052$ & $4.139$ & $2.969\pm 0.472$& $4.171$ \tabularnewline
\hline
$1.5$ & $0.960 \pm 0.079$ &$0.504$ &$0.945\pm 0.073$ & $0.755$ & $0.927\pm 0.076$ & $0.955$ &$0.781 \pm 0.062$ & $3.550$ & $3.371\pm 0.624$& $3.798$ \tabularnewline
\hline
$1.7$ & $0.955 \pm 0.091$ &$0.492$ &$0.919\pm 0.081$ & $0.996$ & $0.931\pm 0.088$ & $0.781$ &$0.751 \pm 0.069$ & $3.612$ & $3.78\pm 0.797$& $3.483$ \tabularnewline
\hline
$1.9$ & $0.943 \pm 0.101$ &$0.557$ &$0.928\pm 0.093$ & $0.770$ & $0.903\pm 0.097$ & $1.003$ &$0.749 \pm 0.079$ & $3.189$ & $3.970\pm 0.942$& $3.152$ \tabularnewline
\hline
\end{tabular}\par\end{centering}
\caption{Null-test \OO$(z)$ with $1\sigma$ errors for the five cosmologies used in this work up to $z=2.0$. We also show the confidence level for each test at each redshifts, values less then $1$ indicate that the null-test is consistence with unity at $1\sigma$, if it is larger it corresponds to the sigmas away the null-test is.
\label{tab:Nulltest-results-binning-mocks-10bins}}
\end{table*}

\section*{Acknowledgments}
The authors thank an anonymous referee for useful suggestions that improved the paper. The authors acknowledge financial support from the Madrid Regional Government (CAM) under the program HEPHACOS S2009/ESP-1473-02, from MICINN under grant FPA2012-39684-C03-02 and Consolider-Ingenio 2010 PAU (CSD2007-00060), as well as from the European Union Marie Curie Initial Training Network UNILHC Granto No. PITN-GA-2009-237920. We also acknowledge the support of the Spanish MINECO's ``Centro de Excelencia Severo Ochoa" Programme under Grant No. SEV-2012-0249.

\begin{appendix}
\section{Null-test for the binning method}

In this section we report the null-test in terms of redshift $z$ that we used for binning the data and its derivatives with respect to four observables to evaluate the propagated error.
The null-test $\mathcal{O}$ becomes:

\begin{widetext}
\be
\mathcal{O}(z)=  \frac{\left(1+z_0\right)^2H(z)}{\left(1+z\right)^2H(z_0)}\frac{f\sigma_{8}(z)}{f\sigma_8(z_0)}\exp\Big\{\frac32 \Omega_{m_0} H_0^2\int_{z_0}^{z}\frac{\left(1+x\right)^2}{H(x)^2f\sigma_8(x)}\left[
\sigma_8(0)\frac{\delta(z_0)}{\delta(z=0)}-\int_{z_0}^{x}\frac{f\sigma_8(y)}{1+y}{\rm d}y\right]{\rm d}x\Big\}
\label{eq:nulltest-z-app}
\ee
\end{widetext}

we chose $z_0$ to be equal the first redshift available, hence all the quantities like $H(z_0)$ and $f\sigma_8(z_0)$ are the first binned values of the data. Eq.~(\ref{eq:nulltest-z-app}) depends also on $H_0$ which is in general a complicated parameter to measure, for this reason we use instead $\Omega_{m_0}H_0^2=100^2\Omega_{m_0}h^2=10^4\omega_{m}$ where $\omega_m$ is a parameter given by Cosmic Microwave Background experiments and easy to measure with great accuracy. It is also important to notice that when $z_0$ approaches to 0 then we have that $H(z_0)\sim H_0$, however, this term should never be thought as the real Hubble constant (like the one appearing in the exponent) but it has to be considered as the value of the Hubble parameter at the lowest redshift because the only true Hubble constant, i.e. that comes directly from the theory is the one appearing in the exponent).

For the sake of completeness we also write the derivatives of Eq.~(\ref{eq:nulltest-z-app}) with respect to the four observables that will be used to evaluate the propagated error on the quantity $\mathcal{O}(z)$ and these are:

\begin{widetext}
\ba
\frac{\partial\log\mathcal{O}(z)}{\partial H(z)} &=& \frac{1}{H(z)}  - 3\times10^4\Omega_{m_0}h^2\int_{0}^{z}\frac{(1+x)^2}{H(x)^3f\sigma_8(x)}\left[\sigma_8(z=0) - \int_{0}^{x}\frac{f\sigma_8(y)}{1+y}{\rm d}y\right]{\rm d}x  \\
~\nn\\
\frac{\partial\log\mathcal{O}(z)}{\partial f\sigma_8(z)} &=& -\frac{1}{f\sigma_8(z)}+\frac32\times10^4\Omega_{m_0}h^2\int_{0}^{z}\frac{(1+x)^2}{H(x)^2f\sigma_8(x)^2}\left[-\sigma_8(z=0) + \int_{0}^{x}\frac{f\sigma_8(y)}{1+y}{\rm d}y-f\sigma_8(x)\log(1+x)\right]{\rm d}x \nn \\
~\\
\frac{\partial\log\mathcal{O}(z)}{\partial \sigma_8(z=0)} &=& \frac32\times10^4\Omega_{m_0}h^2\int_{0}^{z}\frac{(1+x)^2}{H(x)^2f\sigma_8(x)}{\rm d}x  \\
~\nn\\
\frac{\partial\log\mathcal{O}(z)}{\partial \Omega_{m_0}h^2} &=& \frac32\times10^4\Omega_{m_0}h^2\int_{0}^{z}\frac{(1+x)^2}{H(x)^2f\sigma_8(x)}\left[\sigma_8(z=0) - \int_{0}^{x}\frac{f\sigma_8(y)}{1+y}{\rm d}y\right]{\rm d}x\,.
\ea

Then, the final errors on \OO$(z)$ will be given by

\be
\frac{\sigma_{\mathcal{O}(z)}}{\left|\mathcal{O}(z)\right|}=\sqrt{\left(\frac{\partial\log\mathcal{O}(z)}{\partial H(z)}\right)^2\sigma_{H(z)}^2+\left(\frac{\partial\log\mathcal{O}(z)}{\partial f\sigma_8(z)}\right)^2\sigma_{f\sigma_8(z)}^2+\left(\frac{\partial\log\mathcal{O}(z)}{\partial \Omega_{m_0}h^2}\right)^2\sigma_{\Omega_{m_0}h^2}^2+\left(\frac{\partial\log\mathcal{O}(z)}{\partial \sigma_8(z=0)} \right)^2\sigma_{\sigma_8(z=0)}^2}
\ee
\end{widetext}

\end{appendix}

{}

\end{document}